\documentclass[12pt]{article}
\usepackage{slashed}
\usepackage{graphicx}
\usepackage{amssymb}
\usepackage{cite}
\usepackage{mcite}
\usepackage[top=2cm, bottom=2cm, left=2cm, right=2cm]{geometry}

\begin{document}
  \bibliographystyle{h-physrev}
  \vskip -1.5cm
  \begin{flushright}
    MAN/HEP/2007/4
  \end{flushright}
  \begin{center}
    {\LARGE {\bf Resonant CP Violation due to }}\\[0.5cm]
    {\LARGE {\bf Heavy Neutrinos at the LHC}}\\[1cm]
    {\large Simon Bray$\,^a$, Jae Sik Lee$\,^b$ and Apostolos
    Pilaftsis$\,^a$}\\[0.3cm] 
    {$^a$\em School of Physics and Astronomy, University of Manchester,\\
      Manchester M13 9PL, United Kingdom}\\
    {$^b$\em Center for Theoretical Physics, School of Physics,\\
      Seoul National University, Seoul, 151-722, Korea}
  \end{center}  
  \begin{abstract}
The observed  light neutrinos may be  related to the  existence of new
heavy  neutrinos  in the  spectrum  of  the SM.  If  a  pair of  heavy
neutrinos  has nearly degenerate  masses, then  CP violation  from the
interference  between   tree-level  and  self-energy   graphs  can  be
resonantly  enhanced.  We  explore  the possibility  of  observing  CP
asymmetries due  to this mechanism at  the LHC. We consider  a pair of
heavy  neutrinos  $N_{1,2}$ with  masses  ranging from  $100-500\,{\rm
GeV}$ and a  mass-splitting $\Delta m_N=m_{N_2}-m_{N_1}$ comparable to
their   widths   $\Gamma_{N_{1,2}}$.   We   find  that   for   $\Delta
m_N\sim\Gamma_{N_{1,2}}$,  the resulting  CP asymmetries  can  be very
large or even maximal and  therefore, could potentially be observed at
the LHC.
  \end{abstract}

\section{Introduction}
The  observation of  neutrino  oscillations has  established that  the
observed neutrinos  are not  massless and so  the Standard  Model (SM)
must     be    extended    in     order    to     accommodate    these
\cite{Fukuda:1998mi,*Apollonio:1999ae,*Ahmad:2002jz,*Ahn:2002up,
*Eguchi:2002dm}. In  order to  explain why the  neutrinos are  so much
lighter than any of the other fermions, it is common to postulate that
in addition  to the three  observed light neutrinos, there  also exist
partner heavy neutrinos. In  order to avoid very stringent constraints
due  to their  non-observation at  the Large  Electron--Positron (LEP)
collider, these  must have masses greater than  about $100\,{\rm GeV}$
\cite{Achard:2001qv}. However, if they  exist with masses greater than
this,  but  less than  about  $500\,{\rm  GeV}$,  they fall  into  the
category of particles that could be produced for the first time at the
Large                Hadron               Collider               (LHC)
\cite{Pilaftsis:1991ug,Datta:1993nm,*Almeida:2000pz,*Panella:2001wq,
Han:2006ip, delAguila:2006dx}.  Complementary  to such a search, heavy
neutrinos could also be observed at a future linear collider.  Several
studies have  been conducted into  their signals at  the International
Linear                          Collider                         (ILC)
\cite{Gluza:1996bz,*Cvetic:1998vg,*Almeida:2000yx,delAguila:2005mf} as
well  as  possible  alternatives   such  as  an  $e^-\gamma$  collider
\cite{Peressutti:2001ms,*Bray:2005wv}.
  
If neutrinos  with such masses do  exist, then in  general they should
break   lepton-flavour  conservation  and,   if  they   are  Majorana,
lepton-number  ($L$) conservation as  well.  Furthermore,  since their
couplings can be complex, they  can also contribute to CP violation. A
scenario  of  particular interest  is  if two  or  more  of the  heavy
neutrinos         are         quasi-degenerate         in         mass
\cite{Flanz:1994yx,*Flanz:1996fb,*Covi:1996wh}.   In   this  case,  CP
violation  can  be  resonantly  enhanced  such  that  for  appropriate
couplings,   CP   asymmetries   can   be   large   or   even   maximal
\cite{Pilaftsis:1997jf}.   By introducing  flavour  symmetries in  the
singlet neutrino  sector, models  can be built  where quasi-degenerate
heavy  neutrinos with  masses of  order the  electroweak  scale appear
naturally    \cite{Pilaftsis:2003gt,*Pilaftsis:2005rv}\footnote{   For
recent        studies       within        supersymmetric       models,
see~\cite{Garbrecht:2006az,*Branco:2006hz}.}.   These  mostly focus  on
models  of resonant  leptogenesis, which  can be  used to  explain the
Baryon  Asymmetry  of  the   Universe  (BAU).   Observation  of  heavy
neutrinos  at   the  LHC,  and  in  particular,   measurements  of  CP
asymmetries due to them would be a way of testing such models.

Since it  is not  known at  this time whether  neutrinos are  Dirac or
Majorana particles, both possibilities  need to be considered. For the
former,     the    collider    signatures     at    the     LHC    are
lepton-flavour-violating (LFV) processes.  For the latter, in addition
to  these,  lepton-number-violating  (LNV)  processes  could  also  be
observed.   Either of  these types  of processes  should  be virtually
background  free since  they are  forbidden in  the SM  which  is both
lepton-number-conserving     (LNC)    and    lepton-flavour-conserving
(LFC). Higher order processes with  light neutrinos in the final state
could  in principle fake  the signals,  but these  can be  excluded by
suitable  kinematical cuts,  e.g.,  by vetoing  on missing  transverse
momentum ($p_T$) cut \cite{Han:2006ip,delAguila:2006dx}.

CP  violation could  show up  in asymmetries  between  possible signal
final states and their  CP-conjugates. Although the initial $pp$ state
is  not   a  CP-eigenstate,  true  CP-violating   observables  can  be
constructed,   either  by  taking   into  account   the  theoretically
calculable  difference   expected  due  to   the  Parton  Distribution
Functions    (PDF's)   \cite{Pumplin:2002vw,Martin:2002aw},    or   by
considering appropriate  ratios of different processes  such that this
factor drops out.

Observation  of  heavy   neutrinos  at  the  LHC  would   be  a  major
discovery. Less  direct evidence could come from  them contributing to
LFV decays, e.g.~$\mu\to e\gamma$, $\mu\to e$ conversion in nuclei, or
(if Majorana)  neutrinoless double beta decay.  The non-observation of
such processes, along with the excellent agreement of electroweak data
to the SM, places limits on the strength of their couplings.

This  paper  is organised  as  follows:  In  Section \ref{models},  we
describe  extensions  of the  SM  which  include  heavy neutrinos  and
discuss the experimental constraints  on them. Two specific models are
considered,  one  predicting   heavy  Majorana  neutrinos,  the  other
Dirac. Section  \ref{signals} is a short discussion  of the signatures
of  such  particles  at  the  LHC, in  particular,  we  classify  them
according  to  whether  or  not   they  are  LNV.   Next,  in  Section
\ref{propagator},  we   present  the  formalism   for  describing  the
propagator  of   a  system  of  two   coupled  quasi-degenerate  heavy
neutrinos.    This  is  based   on  the   field-theoretic  resummation
approached   developed  in   \cite{Pilaftsis:1997dr}   for  describing
resonant  transitions involving the  mixing of  intermediate fermionic
states.

Assuming the signals  are due entirely to two  nearly degenerate heavy
(Dirac or Majorana) neutrinos, the $2\times 2$ propagator matrices for
such a pair  are then used in Section  \ref{CPV} to derive expressions
for the CP  asymmetries between them. We give  example scenarios where
these asymmetries are large  and discuss their compatibility with both
experimental and theoretical constraints. In Section \ref{results}, we
use these  scenarios in  order to produce  numerical estimates  of the
possible level  of leptonic  CP violation observable  at the  LHC.  We
define a number of CP-violating observables and plot these, along with
the  cross   sections  they   are  derived  from.    Finally,  Section
\ref{conclusion} contains our conclusions.

\setcounter{equation}{0}
\section{Heavy Neutrino Extensions of the Standard Model}\label{models}

  \subsection{Heavy Majorana Neutrino Model}
We first describe the SM, minimally extended to include right-handed
neutrinos. Assuming the Higgs sector is not extended, the Lagrangian
describing the neutrino masses and mixings reads
  \begin{equation}
    \mathcal{L}_{m_\nu}=-\frac{1}{2}\left(\begin{array}{cc}\overline{\nu_L^0}&\overline{{(\nu_R^0)}^c}\end{array}\right)\left(\begin{array}{cc}0&m_D\\m_D^T&m_M\end{array}\right)\left(\begin{array}{c}{(\nu_L^0)}^c\\\nu_R^0\end{array}\right)+{\rm h.c.},
  \end{equation}
where $\nu_L^0$ and  $\nu_R^0$ denote column vectors of  the left- and
right-handed  neutrino fields  in  the weak  basis,  and the  notation
$\nu^c\equiv   C\overline\nu^T$   represents   the  charge   conjugate
fields. Although it is commonly assumed that there is one right-handed
neutrino  per  generation  (as  required  in  $SO(10)$  Grand  Unified
Theories (GUT's) \cite{Fritzsch:1974nn}), this  needs not be so in the
bottom  up approach  considered here.  In fact,  as will  be  shown in
Section \ref{CPV}, in the context  of searches for CP violation, it is
phenomenologically  more interesting  if the  model contains  at least
four  right-handed states. In  order to  maintain generality,  we will
consider  adding $n_R$  right-handed states,  where $n_R$  can  be any
positive integer. The elements of the complex matrices $m_D$ and $m_M$
give  rise  to  Dirac  and  Majorana  mass  terms  for  the  neutrinos
respectively.  The only constraint  on their  structure is  that $m_M$
must be symmetric.
  
The   Majorana   mass-eigenstate   neutrinos   are  related   to   the
weak-eigenstates through
  \begin{equation}
    \left(\begin{array}{cc}\nu_L\\N_L\end{array}\right)=U^T\left(\begin{array}{c}\nu_L^0\\{(\nu_R^0)}^c\end{array}\right).
  \end{equation}
The  states  represented  by   $\nu$  are  the  three  observed  light
neutrinos,  whereas $N$  represents  extra heavy  neutrinos (of  which
there  will be  as many  as right-handed  weak-eigenstates). $U$  is a
$(3+n_R)\times(3+n_R)$ unitary matrix chosen such that
  \begin{equation}
    U^T\left(\begin{array}{cc}0&m_D\\m_D^T&m_M\end{array}\right)U={\rm diag}(m_1,\dots,m_{3+n_R})\,,
    \label{eq1}
  \end{equation}
where $m_1,\dots,m_{3+n_R}$ are the physical neutrino masses.
  
Since $m_D$ is derived from the Higgs mechanism, it is most natural to
assume that  its elements  should be of  order the  vacuum expectation
value of  the Higgs field. By  contrast, $m_M$ is unrelated  to any SM
observables  and  so  could  be  as  large  as  the  GUT  scale.  This
observation leads to the popular seesaw mechanism by which the extreme
smallness of the light neutrino masses are explained through the large
hierarchy                in                these                scales
\cite{Minkowski:1977sc,*Gell-Mann:1979,*Yanagida:1979,*Mohapatra:1979ia}. For
a recent  discussion within  the context of  GUT neutrino  models, see
\cite{Ellis:2006mg}. Unfortunately,  generic seesaw scenarios  are not
phenomenologically  very  interesting since  the  heavy neutrinos  are
predicted to be extremely heavy (of order the GUT scale) and also have
their couplings  to SM  particles highly suppressed.  More interesting
scenarios  for   collider  physics   can  be  formed   by  introducing
approximate  flavour  symmetries that  impose  structure  on the  mass
matrices                 $m_D$                and                $m_M$
\cite{Witten:1985bz,*Mohapatra:1986bd,Pilaftsis:1991ug,Gluza:2002vs,*Altarelli:2004za,Pilaftsis:2003gt,*Pilaftsis:2005rv}. This
can  then  allow  the   heavy  neutrino  couplings  to  be  completely
independent  of the  light neutrino  masses.  In such  theories it  is
possible to have heavy neutrinos with masses of order $100\,{\rm GeV}$
and  significant   couplings  to   SM  particles,  without   being  in
contradiction with light neutrino data.

Writing  the Lagrangian  for  neutrino interactions  in  terms of  the
mass-eigenstates gives \cite{Pilaftsis:1991ug}
  \begin{eqnarray}
    \mathcal{L}_{W^{\pm}}&=&-\frac{g}{\sqrt{2}}\,W_\mu^-\,\overline{l}_i\,\gamma^\mu\,P_L\,B_{ij}\left(\begin{array}{c}\nu\\N\end{array}\right)_j+{\rm h.c.},\label{eq3}\\
    \mathcal{L}_{G^\pm}&=&-\frac{g}{\sqrt{2}M_W}\,G^-\,\overline{l}_i\,\left[m_{l_i}P_L-m_jP_R\right]B_{ij}\left(\begin{array}{c}\nu\\N\end{array}\right)_j+{\rm h.c.},\label{eq4}\\
    \mathcal{L}_{Z}&=&-\frac{g}{4{\rm cos}\theta_w}\,Z_\mu\left(\begin{array}{cc}\overline{\nu}&\overline{N}\end{array}\right)_i\gamma^\mu\left[P_L\,C_{ij}-P_R\,C^*_{ij}\right]\left(\begin{array}{c}\nu\\N\end{array}\right)_j,\label{eq5}\\
    \mathcal{L}_H&=&-\frac{g}{4M_W}\,H\left(\begin{array}{cc}\overline{\nu}&\overline{N}\end{array}\right)_i\left[(m_iP_L+m_jP_R)C_{ij}+(m_jP_L+m_iP_R)C^*_{ij}\right]\left(\begin{array}{c}\nu\\N\end{array}\right)_j,\\
    \mathcal{L}_{G^0}&=&-\frac{ig}{4M_W}\,G^0\left(\begin{array}{cc}\overline{\nu}&\overline{N}\end{array}\right)_i\left[(m_iP_L-m_jP_R)C_{ij}+(m_jP_L-m_iP_R)C^*_{ij}\right]\left(\begin{array}{c}\nu\\N\end{array}\right)_j.\label{eq6}
  \end{eqnarray}
The matrices $B$ and $C$ in the above are given by
  \begin{equation}
    B_{ij}=\sum_{k=1}^3V^*_{Lki}U_{kj}^*\,;\hspace{1cm} C_{ij}=\sum_{k=1}^3U_{ki}U_{kj}^*=\sum_{k=1}^{3+n_R}C_{ik}C^*_{jk}\,,
  \end{equation}
where  $V_L$ is  a  $3\times3$  unitary matrix  relating  the weak  to
mass-eigenstates of  the left-handed charged leptons.  Without loss of
generality, we assume that there  is no mixing in the charged leptons,
i.e.  $V_L$ is  the unit  matrix.  This allows  $B$ to  be written  as
$B_{li}$ with $l=e,\mu,\tau$ and so
  \begin{equation}
    C_{ij}=\sum_{l=1}^3B^*_{li}B_{lj}.
    \label{eq32}
  \end{equation}
  From (\ref{eq1}), the neutrino couplings have to satisfy the constraints
  \begin{equation}
    \sum_{i=1}^{3+n_R}m_iB_{li}B_{l'i}=0\,.
    \label{eq17}
  \end{equation}
These  are important  when it  comes to  considering the  viability of
coupling  scenarios which give  rise to  CP violation,  as is  done in
Section \ref{CPV}.
  
  \subsection{Heavy Dirac Neutrino Model}\label{Dirac}
An alternative model in which  the heavy neutrinos are Dirac particles
can  be constructed  by  adding left-handed  singlets  $S_L^0$ to  the
theory   in    addition   to   the    right-handed   neutrino   fields
$\nu_R^0$. These have no couplings  to SM particles and only enter the
theory through their mixings with the other neutrinos. For simplicity,
we shall  assume that  the same number  of right-handed  neutrinos and
left-handed singlets are  included. This model can be  obtained as the
low  energy limit of  GUT's based  on $SO(10)$  \cite{Wyler:1982dd} or
$E_6$
\cite{Witten:1985bz,*Mohapatra:1986bd,Nandi:1985uh,*Valle:1990pk}
gauge groups.

To obtain a theory with Dirac neutrinos, $B-L$ conservation is imposed
as a global symmetry. The Lagrangian for neutrino masses is then given
by
  \begin{equation}
    \mathcal{L}^{\rm mass}_{\nu}=-\frac{1}{2}\left(\begin{array}{ccc}\overline{\nu_L^0}&\overline{{(\nu_R^0)}^c}&\overline{S_L^0}\end{array}\right)\left(\begin{array}{ccc}0&m_D&0\\m_D^T&0&M^T\\0&M&0\end{array}\right)\left(\begin{array}{c}{(\nu_L^0)}^c\\\nu_R^0\\{(S_L^0)}^c\end{array}\right)+{\rm h.c.}\,.
    \label{eq2}
  \end{equation}
As in  the previous  model, $m_D$ and  $M$ are complex  matrices. This
mass matrix can be diagonalised through the rotations
  \begin{equation}
    \left(\begin{array}{c}\nu_L\\S_L\end{array}\right)=U_L^T\left(\begin{array}{c}\nu_L^0\\S_L^0\end{array}\right);\hspace{1cm}\nu_R=U_R\nu_R^0\,,
  \end{equation}
  where $U_L$ is a $(3+n_R)\times(3+n_R)$ and $U_R$ a $n_R\times n_R$ unitary matrix. If these are chosen appropriately, the Lagrangian given in (\ref{eq2}) can then be expressed as
  \begin{equation}
    \mathcal{L}^{\rm mass}_{\nu}=-\frac{1}{2}\left(\begin{array}{ccc}\overline{\nu_L}&\overline{{(\nu_R)}^c}&\overline{S_L}\end{array}\right)\left(\begin{array}{ccc}0&0&0\\0&0&M_N\\0&M_N&0\end{array}\right)\left(\begin{array}{c}{(\nu_L)}^c\\\nu_R\\{(S_L)}^c\end{array}\right)+{\rm h.c.}\,,
  \end{equation}
with  $M_N$ diagonal.  With  the identifications  $S_L\equiv N_L$  and
$\nu_R\equiv  N_R$,  this is  then  a  mass  term for  three  massless
neutrinos $\nu$\footnote{Although this has already been ruled out, the
model can be made compatible  with experiment by adding small Majorana
mass  terms   for  the  singlets,  e.g.~$\mu\overline{S^0_L}(S^0_L)^c$
\cite{Nandi:1985uh,*Valle:1990pk}.    After   diagonalisation,   these
translate to  small Majorana masses for the  light neutrinos. However,
this  will have  no effect  on any  collider observables,  since these
masses are  tiny compared  to the energy  scales involved.}  and $n_R$
massive Dirac neutrinos $N$.

The three weak-eigenstates $\nu_L^0$ in this theory are related to the
mass-eigenstates through a $(3+n_R)\times(3+n_R)$ unitary matrix, just
like as in the previous theory without singlets. Hence, the Lagrangian
for their  interactions with  $W^\pm$ and $G^\pm$  bosons is  given by
(\ref{eq3})  and (\ref{eq4}),  just with  $U_L$ replacing  $U$  in the
definition  of  $B_{lj}$.   However,  since the  neutrinos  are  Dirac
particles, the Lagrangian for their  couplings to the $Z$, $H$ and
$G^0$ bosons differs from the corresponding one for Majorana particles
[c.f.~(\ref{eq5})--(\ref{eq6})].  It  only contains terms proportional
to $C_{ij}$, not $C^*_{ij}$, and is given by
  \begin{eqnarray}
    \mathcal{L}_{Z}&=&-\frac{g}{2{\rm cos}\theta_w}\,Z_\mu\left(\begin{array}{cc}\overline{\nu}&\overline{N}\end{array}\right)_i\gamma^\mu\,P_L\,C_{ij}\left(\begin{array}{c}\nu\\N\end{array}\right)_j,\\
    \mathcal{L}_{H}&=&-\frac{g}{2M_W}\,H\left(\begin{array}{cc}\overline{\nu}&\overline{N}\end{array}\right)_i\left(m_i\,P_L+m_j\,P_R\right)C_{ij}\left(\begin{array}{c}\nu\\N\end{array}\right)_j,\\
    \mathcal{L}_{G^0}&=&-\frac{ig}{2M_W}\,G^0\left(\begin{array}{cc}\overline{\nu}&\overline{N}\end{array}\right)_i\left(m_i\,P_L-m_j\,P_R\right)C_{ij}\left(\begin{array}{c}\nu\\N\end{array}\right)_j,
  \end{eqnarray}
where again $U_L$ replaces $U$ in the definition of $C_{ij}$.

Dirac  neutrinos can  be considered  as  the limit  of two  degenerate
Majorana neutrinos,  say $N_i$ and $N_j$, whose  couplings are related
through  $B_{li}=iB_{lj}$. It  is easy  to  see then  that for  these,
Eq.~(\ref{eq17}) is automatically satisfied  and hence will not act as
a constraint for Dirac neutrinos.

  \subsection{Constraints on the heavy neutrino couplings}
In both  of the models described above,  the weak-eigenstate neutrinos
are related to the mass-eigenstates through
  \begin{equation}
    \nu_{Ll}^0=\sum_{i=1}^{3+n_R}B_{li}\left(\begin{array}{c}\nu_L\\N_L\end{array}\right)_i,
  \end{equation}
  where $l=e,\mu,\tau$. To make the notation clearer, $B$ is split up into two parts relating to the light and heavy states
  \begin{equation}
    \nu_{Ll}^0=\sum_{i=1}^3B_{l\nu_i}\nu_{Li}+\sum_{i=1}^{n_R}B_{lN_i}N_{Li}.
  \end{equation}
  We can then define the parameter $\Omega_{ll'}$ as \cite{delAguila:2005mf}
  \begin{equation}
    \Omega_{ll'}\equiv\delta_{ll'}-\sum_{i=1}^3B_{l\nu_i}B_{l'\nu_i}^*=\sum_{i=1}^{n_R}B_{lN_i}B_{l'N_i}^*,
  \end{equation}
which  is   a  generalisation  of   the  Langacker--London  parameters
$(s_L^{\nu_{e,\mu,\tau}})^2$    \cite{Langacker:1988ur},    with   the
identification  $\Omega_{ll}=(s_L^{\nu_l})^2$.  The $3\times3$  matrix
$B_{l\nu}$      is,     to      a     good      approximation,     the
Pontecorvo--Maki--Nakagawa--Sakata            (PMNS)            matrix
\cite{Pontecorvo:1957cp,*Pontecorvo:1957qd,*Maki:1962mu},  giving  the
mixing of three left-handed neutrinos. Any deviation from unitarity of
the  PMNS matrix  is  given by  $\Omega_{ll'}$,  and would  constitute
evidence for new physics, such as heavy neutrinos.

Constraints on $\Omega_{ll'}$ come from LEP and low-energy electroweak
  data
  \cite{Langacker:1988ur,Cheng:1980tp,*Korner:1992an,*Bernabeu:1993up,
  *Burgess:1993vc,*Nardi:1994iv,*Bhattacharya:1994bj,*Deppisch:2005zm,
  Ilakovac:1994kj,Bergmann:1998rg,Illana:2000ic,*Cvetic:2002jy}. Tree-level
  processes with light  neutrinos in the final state can  be used as a
  probe  by looking  for a  reduction of  the couplings  of  the light
  neutrinos from their SM values.  A global analysis of such processes
  gives the upper limits \cite{Bergmann:1998rg}
  \begin{equation}
    \Omega_{ee}\le0.012\,;\hspace{1cm}\Omega_{\mu\mu}\le0.0096\,;\hspace{1cm}\Omega_{\tau\tau}\le0.016\,,
    \label{eq20}
  \end{equation}
at  $90\%$ confidence level.  These are  mostly model  independent and
depend only weakly on the heavy neutrino masses.
  
LFV decays  such as $\mu,\tau\to e\gamma$,  $\mu,\tau\to eee$, $\mu\to
e$  conversion  in  nuclei  and  $Z\to  l^+l'^-$  also  constrain  the
couplings.  Heavy neutrinos contribute  in loops,  as such  the limits
obtained from  these depend  on the heavy  neutrino masses  and Yukawa
couplings \cite{Ilakovac:1994kj}.  For $m_N\gg M_W$  and $m_D\ll M_W$,
the   limits  derived   including  recent   analyses  of   Babar  data
\cite{Aubert:2005ye,*Aubert:2005wa}, are
  \begin{equation}
    |\Omega_{e\mu}|\,\lesssim\,0.0001\,;\hspace{1cm}|\Omega_{e\tau}|\,\lesssim\,0.02\,;\hspace{1cm}|\Omega_{\mu\tau}|\,\lesssim\,0.02\,.
    \label{eq30}
  \end{equation}
Since the modulus lies outside of the sums, the applicability of these
limits  to  the  individual  couplings  is limited  as  there  can  be
cancellations   between  the   contributions   from  different   heavy
neutrinos.  Furthermore, lepton-flavour  violation is  a  very general
signature of beyond the SM physics.  Contributions from SUSY particles
for example, could also create cancellations reducing the sum.
  
Attempts to set limits from neutrinoless double beta decay experiments
run into  similar problems. Non-observation translates to  a bound for
Majorana neutrinos of \cite{Belanger:1995nh}
  \begin{equation}
    \left|\sum_i\frac{B^2_{eN_i}}{m_{N_i}}\right|<5\times 10^{-8}\,GeV^{-1}.
  \end{equation}
Although  this  would  appear  to severely  constrain  heavy  Majorana
neutrino couplings to the electron, this is again a sum in which large
cancellations   can  occur   between   contributions  from   different
particles. In  particular, if  heavy neutrinos are  pseudo-Dirac, this
constraint can  be avoided,  as there is  an extra  suppression factor
$(m_{N_2}-m_{N_1})/(m_{N_2}+m_{N_1})$.

\setcounter{equation}{0}
  \section{LHC Signals}\label{signals}
\begin{figure}[ht]
  \begin{center}
    \includegraphics[scale = 2]{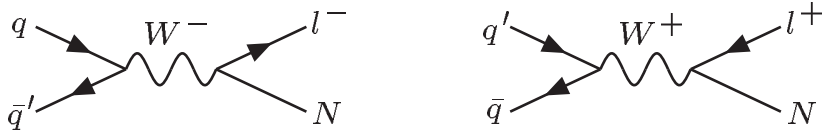}
  \end{center}
  \caption{\it Feynman diagrams for the parton-level subprocesses
  relevant to heavy neutrino production at the LHC.}
  \label{LHC}
\end{figure}

At the LHC, the dominant  production mechanisms for heavy neutrinos if
they have masses in the $100-500\,{\rm GeV}$ range will be $q\overline
q'\to(W^\pm)^*\to    l^\pm   N$,    as    shown   in    Fig.~\ref{LHC}
\cite{Pilaftsis:1991ug,Datta:1993nm,*Almeida:2000pz,*Panella:2001wq,Han:2006ip,delAguila:2006dx}. Due
to  the enhanced contribution  from the  valence quarks,  $W^+$ bosons
will be  produced more copiously  than their charge  conjugates, hence
the  process   with  an  intermediate  $W^+$  will   give  the  larger
signal. Since the Feynman graphs in Fig.~\ref{LHC} have a $W$ boson in
the $s$-channel,  the signal cross  section falls dramatically  as the
mass of  the heavy neutrino is  increased. Hence, even  though the LHC
centre--of--mass--system (cms) energy will be 14 TeV, there is only an
observable signal  for neutrinos  below about $400-500\,{\rm  GeV}$ at
most.

By far the cleanest signals  come from the heavy neutrinos decaying as
$N\to l^\pm W^\mp\to l^\pm jj$,  with $l=e,\mu$ and $j$ representing a
hadronic jet.  All the decay  products can then be  detected, allowing
the reconstruction  of the invariant  mass, and more  importantly, the
observation of lepton-flavour  and lepton-number violation. Given this
decay  chain, signals that  conserve $L$  will be  of the  form $pp\to
l^\pm l'^\mp W^\pm X$, where here $X$ represents the beam remnants and
the $W$ boson is assumed to decay hadronically.  In addition to these,
for Majorana  neutrinos only, LNV  processes of the form  $pp\to l^\pm
l'^\pm W^\mp  X$ are also  possible.  If observed, this  would unravel
the Dirac or Majorana nature of the heavy neutrinos.

In order  to suppress the  SM background, signals for  heavy neutrinos
must  be   LFV  (which  includes   any  LNV  processes).   Since  both
lepton-flavour and  lepton-number violation  are forbidden in  the SM,
the backgrounds  to these processes  require extra light  neutrinos in
the final state. The main source  of this type of background will come
from  three $W$ bosons.  If two  of these  decay leptonically  and the
third  hadronically,  this  can   mimic  the  signal  apart  from  the
additional  light  neutrinos. Recent  analyses  of  this process  have
concluded that  such a  background can be  made negligible  after cuts
\cite{Han:2006ip,delAguila:2006dx}. In particular, a missing $p_T$ cut
is very effective, since this should have no effect on the signal.

\setcounter{equation}{0}
\section{The Resummed Heavy Neutrino Propagator}\label{propagator}

CP violation  may originate from  self-energy, vertex or  higher order
quantum  corrections.  In  general  these  are  small  effects,  since
electroweak loop corrections themselves  are small. However, if two or
more of  the heavy  neutrinos are nearly  degenerate in mass,  then CP
violation from  self-energy corrections (often  termed $\epsilon$-type
CP  violation  \cite{Flanz:1994yx,*Flanz:1996fb,*Covi:1996wh}) can  be
resonantly enhanced \cite{Pilaftsis:1997jf}. In  fact, in the limit of
degenerate  heavy neutrinos,  finite-order perturbation  theory breaks
down.  A  well  defined   field-theoretic  formalism  is  based  on  a
resummation         of          the         self-energy         graphs
\cite{Pilaftsis:1997dr,Pilaftsis:1997jf}. This  approach is manifestly
gauge   invariant   within   the   Pinch  Technique   (PT)   framework
\cite{Cornwall:1989gv,*Papavassiliou:1989zd,*Binosi:2002ft,*Binosi:2003rr,Papavassiliou:1995fq,*Papavassiliou:1995gs,*Papavassiliou:1996zn}
and maintains other field-theoretic  properties, such as unitarity and
CPT  invariance. Our  formalism involves  the absorptive  part  of the
heavy  neutrino self-energy, which  is computed  here at  the one-loop
level. An  important point regarding  this formalism is that  both the
diagonal and off-diagonal elements of the self-energy must be inserted
into the heavy neutrino propagator  matrix. This is crucial, since for
small mass-splittings, the off-diagonal elements play a major role.

Following  this approach,  the propagator  for a  system of  two heavy
neutrinos is given by\footnote{In  all that follows the light neutrino
masses have  been neglected.  Hence, to simplify  the notation  we use
$m_i\equiv m_{N_i}$. Also, we  shall use $B_{li}\equiv B_{lN_i}$, this
should not create any confusion since we are concerned hereafter, only
with the couplings of the heavy neutrinos.}
  \begin{equation}
    \hat{S}(\slashed{p})=\left[\begin{array}{cc}\slashed{p}-m_1+i{\rm Im}\,\hat{\Sigma}_{11}(\slashed{p})&i{\rm Im}\,\hat{\Sigma}_{12}(\slashed{p})\\i{\rm Im}\,\hat{\Sigma}_{21}(\slashed{p})&\slashed{p}-m_2+i{\rm Im}\,\hat{\Sigma}_{22}(\slashed{p})\end{array}\right]^{-1},
    \label{eq7}
  \end{equation}
where  ${\rm  Im}\,\hat{\Sigma}_{ij}(\slashed{p})$  is the  absorptive
part of the heavy neutrino self-energy.

\subsection{Majorana neutrinos}

For Majorana  neutrinos, ${\rm Im}\,\hat{\Sigma}_{ij}(\slashed{p})$ is
of the form
  \begin{equation}
    {\rm Im}\,\hat{\Sigma}_{ij}(\slashed{p})=A_{ij}(s)\slashed{p}P_L+A^*_{ij}(s)\slashed{p}P_R\,,
    \label{eq26}
  \end{equation}
where $s=p^2$ and $A(s)$ is Hermitian, (i.e.~$A_{ij}=A^*_{ji}$).

Writing the propagator as
  \begin{equation}
    \hat{S}_M(\slashed{p})=D_M(s)\slashed{p}P_L+E_M(s)\slashed{p}P_R+F_M(s)P_L+G_M(s)P_R\,,
    \label{eq8}
  \end{equation}
  the matrices $D_M$, $E_M$, $F_M$ and $G_M$ are given by
  \begin{eqnarray}
    D_M&=&\frac{1}{Z_M}\left(\begin{array}{cc}X_{22}\tilde{A}_{11}+s|A_{12}|^2\tilde{A}_{22}&-i(sA_{21}Y+m_1m_2A_{12})\\-i(sA_{12}Y+m_1m_2A_{21})&X_{11}\tilde{A}_{22}+s|A_{12}|^2\tilde{A}_{11}\end{array}\right),\label{eq9}\\
    E_M&=&\frac{1}{Z_M}\left(\begin{array}{cc}X_{22}\tilde{A}_{11}+s|A_{12}|^2\tilde{A}_{22}&-i(sA_{12}Y+m_1m_2A_{21})\\-i(sA_{21}Y+m_1m_2A_{12})&X_{11}\tilde{A}_{22}+s|A_{12}|^2\tilde{A}_{11}\end{array}\right),\\
    F_M&=&\frac{1}{Z_M}\left(\begin{array}{cc}X_{22}m_1-sm_2A^2_{12}&-is(m_1A_{21}\tilde{A}_{22}+m_2A_{12}\tilde{A}_{11})\\-is(m_1A_{21}\tilde{A}_{22}+m_2A_{12}\tilde{A}_{11})&X_{11}m_2-sm_1A^2_{21}\end{array}\right),\hspace{1cm}\\
    G_M&=&\frac{1}{Z_M}\left(\begin{array}{cc}X_{22}m_1-sm_2A^2_{21}&-is(m_1A_{12}\tilde{A}_{22}+m_2A_{21}\tilde{A}_{11})\\-is(m_1A_{12}\tilde{A}_{22}+m_2A_{21}\tilde{A}_{11})&X_{11}m_2-sm_1A^2_{12}\end{array}\right),\label{eq10}
  \end{eqnarray}
  where
  \begin{eqnarray}
    &\tilde{A}_{ii}=1+iA_{ii}\,;\hspace{1cm}X_{ii}=s\tilde{A}^2_{ii}-m^2_{i}\,;\hspace{1cm}Y=|A_{12}|^2+\tilde{A}_{11}\tilde{A}_{22}\,;&\nonumber\\
    &Z_M=X_{11}X_{22}+sm_1m_2(A^2_{12}+A^2_{21})+s^2|A_{12}|^2(2\tilde{A}_{11}\tilde{A}_{22}+|A_{12}|^2)\,.&
    \label{eq27}
  \end{eqnarray}  
  By inspection, it can be seen that $E_M$ is related to $D_M$, and $G_M$ to $F_M$ through
  \begin{equation}
    E_M[A^*]=D_M[A]\,;\hspace{0.8cm}G_M[A^*]=F_M[A]\,;\hspace{0.8cm}E_M=D_M^T\,;\hspace{0.8cm}F_M=F_M^T\,;\hspace{0.8cm}G_M=G_M^T\,.
  \end{equation}
  
  Defining the matrices $M$ and $\tilde{A}$ as
  \begin{equation}
    M=\left(\begin{array}{cc}m_1&0\\0&m_2\end{array}\right);\hspace{1 cm}\tilde{A}=\left(\begin{array}{cc}1+iA_{11}&iA_{12}\\iA_{21}&1+iA_{22}\end{array}\right),
  \end{equation}
  Eq.~(\ref{eq7}) can be written
  \begin{equation}
    \hat{S}_M^{-1}(\slashed{p})=\tilde{A}\slashed{p}P_L+\tilde{A}^T\slashed{p}P_R-M\,.
  \end{equation}
Combining this  with (\ref{eq8}),  the property $SS^{-1}=1$  gives the
further equalities
  \begin{equation}
    sE_M\tilde{A}-F_MM=1\,;\hspace{0.8cm}sD_M\tilde{A}^T-G_MM=1\,;\hspace{0.8cm}G_M\tilde{A}=D_MM\,;\hspace{0.8cm}F_M\tilde{A}^T=E_MM\,.
    \label{eq11}
  \end{equation}
These relations, which can be directly verified for the matrices given
in  (\ref{eq9})--(\ref{eq10})  are  important  for checking  that  CPT
invariance is preserved  in the theory. More details  will be given in
Section~\ref{CPV}.

  \subsection{Dirac neutrinos}
  For Dirac neutrinos, the absorptive part of their self-energy has only the left-handed component, viz
  \begin{equation}
    {\rm Im}\,\hat{\Sigma}_{ij}(\slashed{p})=A_{ij}(s)\slashed{p}P_L\,.
  \end{equation}
  Writing the propagator in the same form as that in (\ref{eq8}), i.e.
  \begin{equation}
    \hat{S}_D(\slashed{p})=D_D(s)\slashed{p}P_L+E_D(s)\slashed{p}P_R+F_D(s)P_L+G_D(s)P_R\,,
  \end{equation}
  the matrices $D_D$, $E_D$, $F_D$ and $G_D$ are given by the expressions
  \begin{eqnarray}
    D_D&=&\frac{1}{Z_D}\left(\begin{array}{cc}(s\tilde{A}_{22}-m_2^2)\tilde{A}_{11}+s|A_{12}|^2&-iA_{12}m_1m_2\\-iA_{21}m_1m_2&(s\tilde{A}_{11}-m_1^2)\tilde{A}_{22}+s|A_{12}|^2\end{array}\right),\\
    E_D&=&\frac{1}{Z_D}\left(\begin{array}{cc}s\tilde{A}_{22}-m_2^2&-isA_{12}\\-isA_{21}&s\tilde{A}_{11}-m_1^2\end{array}\right),\label{eq14}\\
    F_D&=&\frac{1}{Z_D}\left(\begin{array}{cc}(s\tilde{A}_{22}-m_2^2)m_1&-ism_2A_{12}\\-ism_1A_{21}&(s\tilde{A}_{11}-m_1^2)m_2\end{array}\right),\\
    G_D&=&\frac{1}{Z_D}\left(\begin{array}{cc}(s\tilde{A}_{22}-m_2^2)m_1&-ism_1A_{12}\\-ism_2A_{21}&(s\tilde{A}_{11}-m_1^2)m_2\end{array}\right),
  \end{eqnarray}
  with 
  \begin{equation}
    Z_D=(s\tilde{A}_{11}-m_1^2)(s\tilde{A}_{22}-m_2^2)+s^2|A_{12}|^2\,,
  \end{equation}
  and $\tilde{A}_{ii}$ defined in (\ref{eq27}).

  The inverse propagator can be expressed as
  \begin{equation}
    \hat{S}_D^{-1}(\slashed{p})=\tilde{A}\slashed{p}P_L+\slashed{p}P_R-M\,,
  \end{equation}
  so the equivalent of the relations in (\ref{eq11}) for Dirac neutrinos are
  \begin{equation}
    sE_D\tilde{A}-F_DM=1\,;\hspace{1 cm}sD_D-G_DM=1\,;\hspace{1 cm}G_D\tilde{A}=D_DM\,;\hspace{1 cm}F_D=E_DM\,.
    \label{eq13}
  \end{equation}
  
\begin{figure}[ht]
    \begin{center}
      \includegraphics[scale = 0.9]{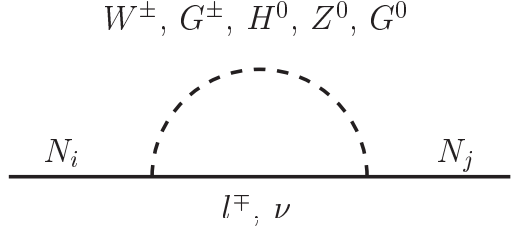}
    \end{center}
    \caption{\it Feynman graphs contributing to the one-loop self-energy of heavy neutrinos. For Dirac neutrinos, only the LNC graphs exist.}
    \label{SelfEnergy}
  \end{figure}

Calculating the  absorptive part of the self-energy  at one-loop level
in the Feynman-'t Hooft gauge\footnote{We  note that the PT result for
fermion self-energies  coincides with that obtained  in the Feynman-'t
Hooft                                                             gauge
\cite{Cornwall:1989gv,*Papavassiliou:1989zd,*Binosi:2002ft,*Binosi:2003rr,Papavassiliou:1995fq,*Papavassiliou:1995gs,*Papavassiliou:1996zn}.}
(for which the Feynman graphs are given in Fig.~\ref{SelfEnergy}), the
matrix  $A(s)$,  which  is  the  same  for  both  Dirac  and  Majorana
neutrinos, is given by
\begin{eqnarray}
  \label{Amatrix} 
   A_{ij}(s) \!&=&\! \frac{g^2}{128\pi
    s^2M_W^2}\Bigg\{\, C_{ij}\, 
\Bigg[\, 4M_W^2(s-M_W^2)^2\theta(s-M_W^2)\: +\:
2M_Z^2(s-M_Z^2)^2\theta(s-M_Z^2)\Bigg]\\
&&\hspace{-1cm} +\:  m_im_j\, C^*_{ij}\, \Bigg[\, 
2(s-M_W^2)^2\theta(s-M_W^2)+(s-M_Z^2)^2\theta(s-M_Z^2)\: +\: 
(s-M_H^2)^2\theta(s-M_H^2)\,\Bigg]\, \Bigg\}\; .\nonumber 
\end{eqnarray}
The  tree-level heavy  neutrino widths  can then  be  obtained through
$\Gamma_{N_i}=m_iA_{ii}(m_i^2)$     for     Dirac    neutrinos     and
$\Gamma_{N_i}=2m_iA_{ii}(m_i^2)$ for Majorana neutrinos.

\setcounter{equation}{0}
  \section{CP Asymmetries in $lW\to l'W$ Type Processes}\label{CPV}
For  the LHC  signals described  in Section  \ref{signals},  the heavy
neutrino propagator  is coupled to a  charged lepton and  $W$ boson at
each   end.  The   asymmetries  between   such  processes   and  their
CP-conjugates  will thus  be the  same  as for  the $2\to2$  processes
$lW\to l'W$. This can be  understood since the fermion line containing
the heavy  neutrino is  the same  in the Feynman  graphs for  both the
$2\to3$ signals and the corresponding $2\to2$ processes. The fact that
one of the  charged leptons changes from being  a final state particle
to an initial  state particle will not effect much the  size of the CP
asymmetries.

As a  result of CPT invariance,  if all possible final  states $X$ are
summed over, then \cite{Pilaftsis:1997dr}
  \begin{equation}
    \sigma(l^+ W^-\to N\to X)=\sigma(l^- W^+\to N\to X).
    \label{eq12}
  \end{equation}
  However, it is possible for the asymmetry between the cross sections for producing any particular final state and its CP-conjugate to be large. 

Considering LNC  processes first, the  CP-violating difference between
$\sigma(q\bar q'\to l^+l'^-W^+)$  and $\sigma(\bar qq'\to l^-l'^+W^-)$
will thus be proportional  to that between $\sigma(l^-W^+\to l'^-W^+)$
and $\sigma(l^+W^-\to l'^+W^-)$. The  only relevant parts of the cross
sections are those that involve  the couplings of the heavy neutrinos,
any    pair   of    CP-conjugate   processes    will    be   otherwise
identical.  Therefore,  this asymmetry  will  be  proportional to  the
factors
  \begin{eqnarray}
    \Delta_{\rm CP}^{\rm LNC}|_{\rm Dirac}&=&\left|B_{l'}E_D(s)B_l^\dagger\right|^2-\left|B_lE_D(s)B_{l'}^\dagger\right|^2\,,\\
    \Delta_{\rm CP}^{\rm LNC}|_{\rm Majorana}&=&\left|B_{l'}E_M(s)B_l^\dagger\right|^2-\left|B_lE_M(s)B_{l'}^\dagger\right|^2\,,
  \end{eqnarray}
depending  on  whether the  heavy  neutrino  is  a Dirac  or  Majorana
particle.  In the  above, $B_l=(B_{l1},B_{l2})$  with $N_1$  and $N_2$
being  the two  nearly degenerate  heavy neutrinos  involved. As  is a
direct consequence  of CPT invariance,  these vanish if $l$  and $l'$,
the two  charged leptons that the  heavy neutrinos couple  to, are the
same. Since $E[B^*]=E^T[B]$ for both Dirac and Majorana neutrinos, the
two  terms in $\Delta_{\rm  CP}^{\rm LNC}$  transform into  each other
under   $B\to    B^*$,   hence   confirming    that   they   represent
CP-conjugates. Using  the expression for $E_D$  given in (\ref{eq14}),
$\Delta_{\rm CP}^{\rm LNC}|_{\rm Dirac}$ is given by
  \begin{eqnarray}
    \Delta_{\rm CP}^{\rm LNC}|_{\rm Dirac}&=&\frac{4s(s-m_2^2)}{|Z_D|^2}\left(A_{11}{\rm Im}[B^*_{l1}B_{l'1}B_{l2}B^*_{l'2}]\right.\nonumber\\
    &&\left.-|B_{l1}|^2{\rm Im}[A_{12}B_{l'1}B^*_{l'2}]+|B_{l'1}|^2{\rm Im}[A_{12}B_{l1}B^*_{l2}]\right)\nonumber\\
    &&+\,(1\leftrightarrow2)\,.
    \label{eq15}
  \end{eqnarray}
The full expression for $\Delta_{\rm CP}^{\rm LNC}|_{\rm Majorana}$ is
rather  more  complicated,  it  is  thus pragmatic  to  work  with  an
approximation when  performing analytic calculations.  The elements of
$A$  are very small  (of order  the heavy  neutrino widths  divided by
their masses), so terms above first  order in them can be dropped to a
very good approximation. Doing  this, the matrix $E_M$ is approximated
as
  \begin{equation}
    E_M\approx\frac{1}{Z_M}\left(\begin{array}{cc}(s-m_2^2)\tilde A_{11}+2isA_{22}&-i(sA_{12}+m_1m_2A_{21})\\-i(sA_{21}+m_1m_2A_{12})&(s-m_1^2)\tilde A_{22}+2isA_{11}\end{array}\right).
  \end{equation}
  Using this approximation, $\Delta_{\rm CP}^{\rm LNC}|_{\rm Majorana}$ is given by
  \begin{eqnarray}
    \Delta_{\rm CP}^{\rm LNC}|_{\rm Majorana}&=&\frac{4(s-m_2^2)}{|Z_M|^2}\left[(s+m_1^2)A_{11}{\rm Im}[B^*_{l1}B_{l'1}B_{l2}B^*_{l'2}]\right.\nonumber\\
      &&-|B_{l1}|^2(s{\rm Im}[A_{12}B_{l'1}B^*_{l'2}]+m_1m_2{\rm Im}[A_{21}B_{l'1}B^*_{l'2}])\nonumber\\
      &&\left.+|B_{l'1}|^2(s{\rm Im}[A_{12}B_{l1}B^*_{l2}]+m_1m_2{\rm Im}[A_{21}B_{l1}B^*_{l2}])\right]\nonumber\\
    &&+\,(1\leftrightarrow2)\,.
    \label{eq16}
  \end{eqnarray}
The propagator  for Dirac  neutrinos simply contains  a subset  of the
terms in  the corresponding  one for Majorana  neutrinos, the  same is
thus  also  true  of  the CP-violating  expressions  (\ref{eq15})  and
(\ref{eq16}). The  extra terms that appear for  Majorana neutrinos all
have an $m^2$  mass dependence, these are due  to interference between
graphs  without a  chirality flip  in them  and graphs  with  a double
chirality flip,  the latter not  appearing for Dirac  neutrinos. There
will also be higher order (in  the elements of $A$) terms for Majorana
neutrinos that have been neglected in our approximation for $E_M$.

An analogous expression  can be derived for the  LNV signals (assuming
neutrinos are Majorana particles so such processes are allowed), which
are  of  the  form  $\sigma(q\bar  q'\to  l^\pm  l'^\pm  W^\mp)$.  The
asymmetries  between  these  signals  are  related  to  those  between
$\sigma(l^-W^+\to l'^+W^-)$ and $\sigma(l^+W^-\to l'^-W^+)$, which are
proportional to
  \begin{equation}
    \Delta_{\rm CP}^{\rm LNV}=\left|B^*_{l'}G_M(s)B_l^\dagger\right|^2-\left|B_lF_M(s)B_{l'}^T\right|^2\,,
  \end{equation}
where only  the contributions  from the resonant  $s$-channel diagrams
are included.  $F_M(s)$ and $G_M(s)$  are symmetric, so both  terms in
this  expression are  individually  invariant under  $l\leftrightarrow
l'$.  Also, since $G_M[B^*]=F_M[B]$,  it can  again be  confirmed that
these terms  represent CP-conjugates,  since they transform  into each
other under $B\to B^*$.

In order to make the  expressions manageable, we will continue to work
with an  approximation of the heavy Majorana  neutrino propagator that
drops higher order terms in the  elements of the matrix $A$. With this
approximation, the matrices $F_M$ and $G_M$ are given by
  \begin{eqnarray}
    F_M&\approx&\frac{1}{Z_M}\left(\begin{array}{cc}m_1(s(1+2iA_{22})-m^2_2)&-is(m_1A_{21}+m_2A_{12})\\-is(m_1A_{21}+m_2A_{12})&m_2(s(1+2iA_{11})-m^2_1)\end{array}\right),\\
    G_M&\approx&\frac{1}{Z_M}\left(\begin{array}{cc}m_1(s(1+2iA_{22})-m^2_2)&-is(m_1A_{12}+m_2A_{21})\\-is(m_1A_{12}+m_2A_{21})&m_2(s(1+2iA_{11})-m^2_1)\end{array}\right).
  \end{eqnarray}
Using these  expressions, the  CP asymmetry for  LNV processes  of the
type considered is
  \begin{eqnarray}
    \Delta_{\rm CP}^{\rm LNV}&=&\frac{4sm_1(s-m_2^2)}{|Z_M|^2}\bigg[2m_2A_{11}{\rm Im}[B_{l1}^*B_{l'1}^*B_{l2}B_{l'2}]\nonumber\\
      &&+|B_{l1}|^2\left(m_1{\rm Im}[A_{12}B_{l'1}B^*_{l'2}]+m_2{\rm Im}[A_{21}B_{l'1}B^*_{l'2}]\right)\nonumber\\
      &&+|B_{l'1}|^2\left(m_1{\rm Im}[A_{12}B_{l1}B^*_{l2}]+m_2{\rm Im}[A_{21}B_{l1}B^*_{l2}]\right)\bigg]\nonumber\\
    &&+\,(1\leftrightarrow2)\,.
    \label{eq19}
  \end{eqnarray}
Since  all first  order graphs  contributing to  LNV processes  have a
single chirality  flip in the propagator,  all the terms  in the above
expression are proportional to $m^2$.

\subsection{Theoretical constraints (for Majorana neutrinos)}\label{3Mconsts}
For  three  heavy  Majorana  neutrinos,  $B_{lN}$  is  a  $3\times  3$
matrix.  Ignoring  the  light  neutrino  masses,  the  constraints  in
(\ref{eq17}) thus leave  four of the heavy neutrino  couplings as free
parameters. For example, $B_{l1}$ ($l=e,\mu,\tau$) and $B_{e2}$ can be
chosen, Eq.~(\ref{eq17}) is then satisfied by
  \begin{equation}
    B_{e3}=\pm i\sqrt{\frac{m_1B_{e1}^2+m_2B_{e2}^2}{m_3}}\,;\hspace{1 cm}B_{li}=\frac{B_{l1}B_{ei}}{B_{e1}}\,.
    \label{eq18}
  \end{equation}
In order to see how  this effects the expressions for the $\Delta_{\rm
CP}$'s,  we   write  the  absorptive  self-energy   matrix  $A$  given
in~(\ref{Amatrix}) as follows:
  \begin{equation}
A_{ij}\ =\ C_{ij}\, a(s)\: +\: C^*_{ij}\, \frac{m_im_j}{M_W^2}\;
b(s)\; ,
  \end{equation}
where $a$ and $b$ are dimensionless real functions. Then, using (\ref{eq32})
  \begin{equation}
{\rm Im}[A_{ij}B_{l1}B^*_{l2}]\ =\ 
\sum_{l'}\: \Bigg( a\,
{\rm Im}[B_{l'i}^*B_{l'j}B_{l1}B^*_{l2}]\: +\: b\,
  \frac{m_im_j}{M_W^2}\, {\rm Im}[B_{l'i}B^*_{l'j}B_{l1}B^*_{l2}]\,
  \Bigg)\; ,
\end{equation}
which, after the application of (\ref{eq18}), becomes
  \begin{equation}
{\rm Im}[A_{ij}B_{l1}B^*_{l2}]\ =\ 
C_{11}\frac{|B_{l1}|^2}{|B_{e1}|^4}
\left(a\, {\rm
  Im}\left[B_{ei}^*B_{ej}B_{e1}B^*_{e2}\right]\: 
+\: b\frac{m_im_j}{M_W^2}\; {\rm
  Im}\left[B_{ei} B^*_{ej}B_{e1}B^*_{e2}\right]\,\right)\; . 
  \end{equation}
Inserting  this into  (\ref{eq16}),  $\Delta_{\rm CP}^{\rm  LNC}|_{\rm
Majorana}$ can  be shown to  vanish, while in (\ref{eq19}),  all terms
proportional to $a$ cancel out and we are left with
  \begin{equation}
    \Delta_{\rm CP}^{\rm LNV}=\frac{8bsm_1m_2(m_1-m_2)|B_{l1}|^2|B_{l'1}|^2C_{11}}{M_W^2|Z_M|^2}\left(\frac{|B_{e2}|^2}{|B_{e1}|^2}m_2(s-m_1^2)+m_1(s-m_2^2)\right){\rm Im}\left[\left(\frac{B_{e2}}{B_{e1}}\right)^2\right],
  \end{equation}
  where
  \begin{equation}
    b=\frac{g^2(2(s-M_W^2)^2\theta(s-M_Z^2)+(s-M_Z^2)^2\theta(s-M_Z^2)+(s-M_H^2)^2\theta(s-M_H^2))}{128\pi s^2}\,.
  \end{equation}
For neutrino masses accessible to colliders this is always very small,
so no  observable CP  violation is possible  for the  model considered
with three heavy Majorana neutrinos. This result holds as long as only
two of them are close enough in mass to have significant mixing in the
propagator. The case of all three neutrinos being nearly degenerate is
more involved, so whether large  CP asymmetries can occur in this case
may be studied elsewhere.

If  at  least four  heavy  neutrinos  exist,  Eq.~(\ref{eq17}) can  be
satisfied for any choices of the couplings of the two quasi-degenerate
neutrinos.  Scenarios which result  in large  CP asymmetries  are thus
possible,  and  it is  in  the  context of  such  a  theory which  our
numerical  results for  Majorana neutrinos  are to  be  considered. As
mentioned  in Section  \ref{Dirac}, Eq.~(\ref{eq17})  is automatically
satisfied for the model with  Dirac neutrinos. As such, even with just
two heavy neutrinos, large CP asymmetries can result here.

  \subsection{Scenarios with large CP asymmetries}\label{large}
For  the  purposes  of  our  numerical  calculations,  we  assume  the
couplings   of  the   nearly  degenerate   neutrinos  can   be  chosen
independently  for both  Dirac and  Majorana neutrinos.  Although this
implies at  least four heavy neutrinos  in the Majorana  case, we also
assume that only two of these are  close in mass, such that we can use
our $2\times2$ propagators given in Section \ref{propagator}. In order
to determine scenarios which would  result in large CP asymmetries, we
consider          just           the          kinematic          point
$s=\overline{s}=\frac{1}{2}(m_1^2+m_2^2)$.  To  further  simplify  the
expressions, we will also set the heavy neutrino couplings to all have
a common magnitude $|B|$.

Introducing            the            notation            $\Delta_{\rm
CP}(\overline{s})\equiv\overline\Delta_{\rm                       CP}$,
$Z(\overline{s})\equiv\overline                 Z$                 and
$A_{ij}(\overline{s})\equiv\overline{A}_{ij}$,  the  CP asymmetry  for
Dirac neutrinos is given by
  \begin{eqnarray}
    \overline\Delta_{\rm CP}^{\rm LNC}|_{\rm Dirac}&=&\frac{m_1^4-m_2^4}{|\overline Z_D|^2}\left[(\overline{A}_{11}+\overline{A}_{22}){\rm Im}[B^*_{l1}B_{l'1}B_{l2}B^*_{l'2}]-\right.\nonumber\\
      &&\left.(|B_{l1}|^2+|B_{l2}|^2){\rm Im}[\overline{A}_{12}B_{l'1}B^*_{l'2}]+(|B_{l'1}|^2+|B_{l'2}|^2){\rm Im}[\overline{A}_{12}B_{l1}B^*_{l2}]\right].
    \label{eq28}
  \end{eqnarray}
One way  to obtain a  large value for  this expression is  to maximise
${\rm  Im}[B^*_{l1}B_{l'1}B_{l2}B^*_{l'2}]$,  which  can  be  done  by
having  one of the  four couplings  imaginary and  the rest  real. The
couplings to the third charged  lepton (i.e.~not $l$ or $l'$) can then
be  chosen such  that $C_{12}$  is either  real or  imaginary,  and so
either   ${\rm    Im}[\overline{A}_{12}B_{l'1}B^*_{l'2}]$   or   ${\rm
Im}[\overline{A}_{12}B_{l1}B^*_{l2}]$ will  be zero. If  we now impose
the condition  that all the  couplings have the same  magnitude $|B|$,
Eq.~(\ref{eq28}) becomes
  \begin{eqnarray}
    \left|\overline\Delta_{\rm CP}^{\rm LNC}|_{\rm Dirac}\right|&=&\frac{|B|^4}{|\overline Z_D|^2}(\overline{A}_{11}+\overline{A}_{22}-2|\overline{A}_{12}|)|m_1^4-m_2^4|\,.
  \end{eqnarray}
  Although there is a partial cancellation between the elements of $\overline{A}$, the off-diagonal element will only be about one third of the magnitude of the diagonal elements (which are approximately equal). This is because in $\overline{A}_{12}$, the contribution from the couplings to the third charged lepton cancel exactly with the contribution from either $l$ or $l'$. The equality $\overline{A}_{11}=\overline{A}_{22}=3|\overline{A}_{12}|$ is not exact however, since there are small differences due to the mass-splitting of the two heavy neutrinos.
  
For  Majorana  neutrinos,  the  expression  for  $\overline\Delta_{\rm
CP}^{\rm LNC}|_{\rm Majorana}$ is
  \begin{eqnarray}
    \overline\Delta_{\rm CP}^{\rm LNC}|_{\rm Majorana}&=&\frac{m_1^2-m_2^2}{|\overline Z_M|^2}\left[\left((3m_1^2+m_2^2)\overline{A}_{11}+(m_1^2+3m_2^2)\overline{A}_{22}\right){\rm Im}[B^*_{l1}B_{l'1}B_{l2}B^*_{l'2}]\right.\nonumber\\
      &&-(|B_{l1}|^2+|B_{l2}|^2)\left((m_1^2+m_2^2){\rm Im}[\overline{A}_{12}B_{l'1}B^*_{l'2}]+2m_1m_2{\rm Im}[\overline{A}_{21}B_{l'1}B^*_{l'2}]\right)\nonumber\\
      &&\left.+(|B_{l'1}|^2+|B_{l'2}|^2)\left((m_1^2+m_2^2){\rm Im}[\overline{A}_{12}B_{l1}B^*_{l2}]+2m_1m_2{\rm Im}[\overline{A}_{21}B_{l1}B^*_{l2}]\right)\right].\hspace{0.2cm}
  \end{eqnarray}
If      $\overline{A}_{12}$     is      imaginary,      then     ${\rm
Im}[\overline{A}_{12}B_{l1}B^*_{l2}]=-{\rm
Im}[\overline{A}_{21}B_{l1}B^*_{l2}]$.  Under the same  assumption for
the $B$-couplings, we find
  \begin{eqnarray}
    \left|\overline\Delta_{\rm CP}^{\rm LNC}|_{\rm Majorana}\right|&=&\frac{|m_1^2-m_2^2||B|^4}{|\overline Z_M|^2}\left((3m_1^2+m_2^2)\overline{A}_{11}+(m_1^2+3m_2^2)\overline{A}_{22}\right)+\mathcal{O}\left[(m_1-m_2)^2\right].\hspace{0.2cm}
  \end{eqnarray}
  
For LNV processes, $\overline\Delta_{\rm CP}^{\rm LNV}$ is given by
  \newpage
  \begin{eqnarray}
    \overline\Delta_{\rm CP}^{\rm LNV}&=&\frac{m_1^4-m_2^4}{|\overline Z_M|^2}\bigg[2m_1m_2(\overline{A}_{11}+\overline{A}_{22}){\rm Im}[B_{l1}^*B_{l'1}^*B_{l2}B_{l'2}]\nonumber\\
      &&+\left((m_1^2|B_{l1}|^2+m_2^2|B_{l2}|^2){\rm Im}[\overline{A}_{12}B_{l'1}B^*_{l'2}]+m_1m_2(|B_{l1}|^2+|B_{l2}|^2){\rm Im}[\overline{A}_{21}B_{l'1}B^*_{l'2}]\right)\nonumber\\
      &&+\left((m_1^2|B_{l'1}|^2+m_2^2|B_{l'2}|^2){\rm Im}[\overline{A}_{12}B_{l1}B^*_{l2}]+m_1m_2(|B_{l'1}|^2+|B_{l'2}|^2){\rm Im}[\overline{A}_{21}B_{l1}B^*_{l2}]\right)\bigg].\hspace{0.6cm}
  \end{eqnarray}
As long as $l\ne l'$, then the same assumptions for the couplings lead to
  \begin{eqnarray}
    \left|\overline\Delta_{\rm CP}^{\rm LNV}\right|&=&\frac{2m_1m_2|B|^4}{|\overline Z_M|^2}(\overline{A}_{11}+\overline{A}_{22})|m_1^4-m_2^4|+\mathcal{O}\left[(m_1-m_2)^2\right].
  \end{eqnarray}
  If $l=l'$, then
  \begin{eqnarray}
    \overline\Delta_{\rm CP}^{\rm LNV}&=&\frac{m_1^4-m_2^4}{|\overline Z_M|^2}\bigg[2m_1m_2(\overline{A}_{11}+\overline{A}_{22}){\rm Im}[(B_{l1}^*B_{l2})^2]\nonumber\\
      &&+2\left((m_1^2|B_{l1}|^2+m_2^2|B_{l2}|^2){\rm Im}[\overline{A}_{12}B_{l1}B^*_{l2}]+m_1m_2(|B_{l1}|^2+|B_{l2}|^2){\rm Im}[\overline{A}_{21}B_{l1}B^*_{l2}]\right).\hspace{0.3cm}
      \label{eq29}
  \end{eqnarray}
The  easiest  way   to  get  a  large  value  for   this  is  to  have
$B_{l1}B^*_{l2}$  imaginary.  ${\rm  Im}[(B_{l1}^*B_{l2})^2]$ is  then
zero, but  both ${\rm Im}[\overline{A}_{12}B_{l1}B^*_{l2}]$  and ${\rm
Im}[\overline{A}_{21}B_{l1}B^*_{l2}]$ will  be large (and  of the same
sign),  as long  as $C_{12}$  has  a significant  real component.  The
asymmetry is then
  \begin{eqnarray}
    \left|\overline\Delta_{\rm CP}^{\rm LNV}\right|&=&\frac{2|B|^4}{|\overline Z_M|^2}|m_1^4-m_2^4|(m_1+m_2)^2{\rm Re}[\overline{A}_{12}]\,.
  \end{eqnarray}
  
In  all   the  above  coupling   scenarios,  if  the   heavy  neutrino
mass-splitting $\Delta  m_N\equiv m_2-m_1$  is of order  their widths,
then  the  CP   asymmetries  can  be  of  order   the  cross  sections
involved.  In the resonant  region, these  have terms  proportional to
either ${\Delta m}^2$, $\Gamma_N^2$  or $\Delta m\,\Gamma_N$ (all over
$|\overline   Z_{D/M}|^2$).   The   CP   asymmetries  are   themselves
proportional to  $\Delta m\,\Gamma_N$ (over  $|\overline Z_{D/M}|^2$),
so for  $\Delta m\approx\Gamma_N$, CP violation can  be resonant. This
still  requires   particular  coupling  scenarios  such   as  we  have
described,  but   since  we   are  interested  in   investigating  the
possibility  of large CP  asymmetries at  the LHC,  we will  use these
optimised conditions for our numerical studies.

\setcounter{equation}{0}
  \section{Numerical Results}\label{results}
In this section  we give our numerical results  for the cross sections
of the various signals of the type discussed in Section \ref{signals},
for clarity, we list these in Table \ref{Tab}. As mentioned in Section
\ref{signals}, in  all signals considered, the $W$  bosons are assumed
to decay hadronically.  We do not consider $\tau$  leptons as possible
final state  particles since these decay before  reaching the detector
and so would require a  more involved analysis. Since we are primarily
interested  in CP asymmetries  between the  signal processes,  we also
plot these, which we define in  later this section and also include in
Table \ref{Tab}. For the results  shown, we have used the CTEQ6M PDF's
\cite{Pumplin:2002vw},  in which  we  have used  $Q=m_N$. Setting  $Q$
equal  to  the invariant  mass  of  the  $W^*$ instead,  for  example,
increases  the  cross  sections  slightly  for  relatively  low  heavy
neutrino masses.

The basic assumption required for our calculations to be valid is that
the  two quasi-degenerate  heavy neutrinos  are the  only  new physics
particles  that  make an  appreciable  contribution  to the  processes
considered. All  others are assumed  to either not couple  to relevant
particles,  or have  masses $\gtrsim1\,{\rm  TeV}$.   Our calculations
depend weakly on  the mass of the Higgs boson, for  which we have used
$M_H=120\,{\rm  GeV}$,   and  we   have  globally  applied   the  cuts
$p_T>15\,{\rm GeV}$  and $|\eta|>2.5$  for all final  state particles.
Note  that these  kinematical  cuts  are similar  to  those chosen  in
\cite{Han:2006ip}.

  We consider the following three scenarios:
  \begin{description}
  \item[{\bf (S1):}] $B_{lN}=0.05$ for $l=e,\mu,\tau$ and $N=1,2$.
  \item[{\bf (S2):}] As {\bf (S1)}, except $B_{e2}=0.05i$.
  \item[{\bf (S3):}] As {\bf (S1)}, except $B_{\mu2}=0.05i$ and $B_{\tau2}=-0.05$.
  \end{description}
The first scenario, where all couplings are real, is the CP-conserving
limit.  This enables  us  to  separate true  CP  asymmetries from  the
asymmetry due to the PDF's.  The second two are chosen specifically as
examples  of resonant  CP  violation,  as guided  by  our analysis  in
Section  \ref{large}.   Scenario  {\bf  (S2)}  gives   rise  to  large
asymmetries  for  Majorana neutrinos,  but  not  for Dirac  neutrinos,
whereas in  scenario {\bf  (S3)} large CP  asymmetries occur  for both
Dirac and Majorana neutrinos, but  only if the two final state charged
leptons are of different flavour.
  
Including just the  two heavy neutrinos that are  directly involved in
the  signal  processes, we  have  $\sum_i|B_{lN_i}|^2=0.005$ (for  all
$l$).  This leaves room for couplings to other heavy neutrinos without
invalidating  the   experimental  limits  in   (\ref{eq20}).  This  is
especially important for Majorana neutrinos, since as noted in Section
\ref{3Mconsts},  it is  necessary then  to  have at  least four  heavy
neutrinos to avoid theoretical  constraints on their couplings. In the
same  context, a  degree of  cancellation between  the  different loop
contributions might  be necessary  for the scenarios  {\bf (S1)}--{\bf
(S3)} to satisfy the  bounds in (\ref{eq30}), especially those derived
from the non-observation of $\mu\to e\gamma$.

\begin{figure}[ht]
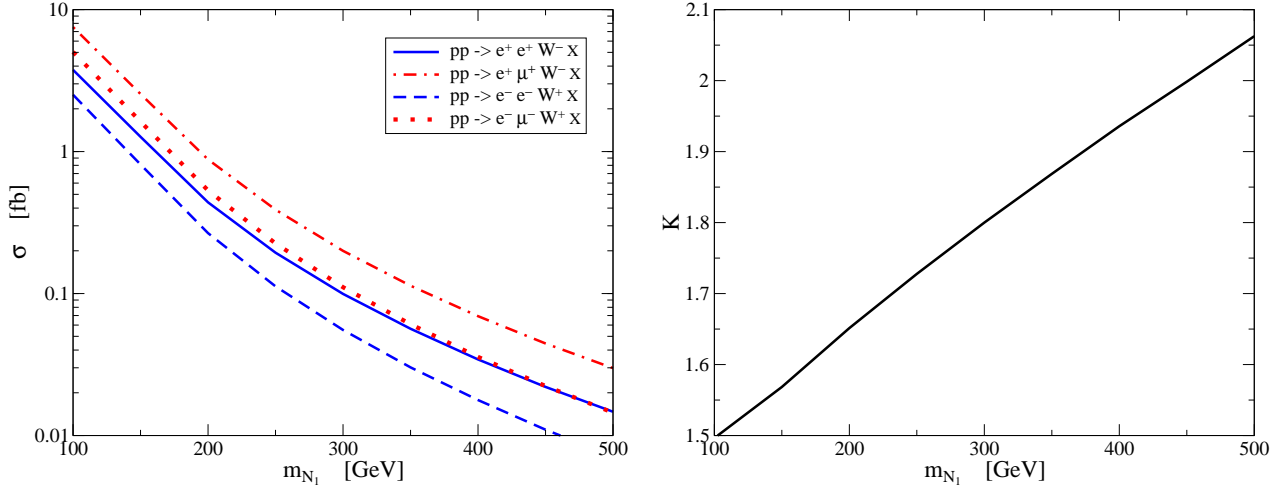

    \vspace{0.2cm}
    \begin{center}
      \includegraphics[scale=0.36]{S1.eps}\hspace{0.3cm}
      \includegraphics[scale=0.36]{K.eps}
    \end{center}
    \caption{\it Left plot: Signal cross sections for scenario {\bf
    (S1)}. Right plot: The function $K$ as defined in
    (\ref{eq21}). All additional signals of the type considered have
    the same cross section as one of the four shown}
    \label{Kplot}
  \end{figure}

We  start  our  investigation  of  CP violation  by  constructing  the
following CP asymmetry
  \begin{equation}
    A_{\rm CP}=\frac{\sigma(pp\to(W^+)^*\to S_i)-K\sigma(pp\to(W^-)^*\to\overline S_i)}{\sigma(pp\to(W^+)^*\to S_i)+K\sigma(pp\to(W^-)^*\to\overline S_i)}\,,
    \label{eq22}
  \end{equation}
where $S_i$  can stand for any  of the signal  final states considered
and $\overline  S_i$ its CP-conjugate. The function  $K$ takes account
of the different  PDF's involved in producing $W^+$  and $W^-$ bosons,
such that $A_{\rm  CP}=0$ if CP is conserved. It  has to be calculated
theoretically and is defined as
  \begin{equation}
    K=\frac{\sigma(pp\to(W^+)^*\to S_i)}{\sigma(pp\to(W^-)^*\to \overline S_i)}\Bigg|_{\slashed{CP}=0}\,,
    \label{eq21}
  \end{equation}
where again $S_i$  can be any of the  possible final states considered
and  $\overline S_i$  its  CP-conjugate. An  important  point is  that
although $A_{\rm CP}$ is in general different for different $S_i$, $K$
is, to a  very good approximation, universal whichever  signal (of the
type considered)  is used  to define it.  Therefore, we  calculate the
cross  sections in the  CP-conserving limit  with all  couplings real,
like for  example in  scenario {\bf (S1)}.  $K$ is independent  of the
magnitudes  of the  couplings and  mass-splitting (for  $\Delta m_N\ll
m_N$),  so  any  other  CP-conserving  scenario would  give  the  same
value.  It is  plotted as  a function  of $m_{N_1}$  (the mass  of the
lighter of the two heavy neutrinos) in Fig.~\ref{Kplot} along with the
signal cross sections in {\bf (S1)}.  The value of $K$ can be obtained
by taking the ratio of either of the two pairs of CP-conjugate signals
shown. In  fact, in  this special scenario,  where all  heavy neutrino
couplings  are equal, all  additional signals  of the  type considered
have the same  cross section as one of the  four shown. However, there
is an element of theoretical uncertainty predominantly coming from the
different  choices of  PDF's used.  For instance,  using  the MRST2004
PDF's \cite{Martin:2002aw} instead, we find values for $K$ that differ
by less  than $1\%$.  They are also  insensitive to  the factorisation
scale $Q$ that is used.

 Another  set  of  CP-violating  observables  can  be  constructed  by
 considering ratios of different  processes, such that the asymmetries
 due to the different PDF's  cancel out. Defining the ratios $R_+$ and
 $R_-$ as
  \begin{equation}
    R_+=\frac{\sigma(pp\to(W^+)^*\to S_i)}{\sigma(pp\to(W^+)^*\to S_j)}\,;\hspace{1cm}R_-=\frac{\sigma(pp\to(W^-)^*\to\overline S_i)}{\sigma(pp\to(W^-)^*\to\overline S_j)}\,,
  \end{equation}
  CP-violating observables can be constructed from
  \begin{equation}
    R_{\rm CP}=\frac{R_+-R_-}{R_++R_-}\,.
    \label{eq23}
  \end{equation}
This  second method  has  the advantage  that  it does  not depend  on
$K$. However,  it is more  complicated to construct and  analyse since
two different  processes (plus  their CP-conjugates) are  required for
each asymmetry.

  \subsection{LNC processes}\label{LNC}
  \begin{figure}[ht!]
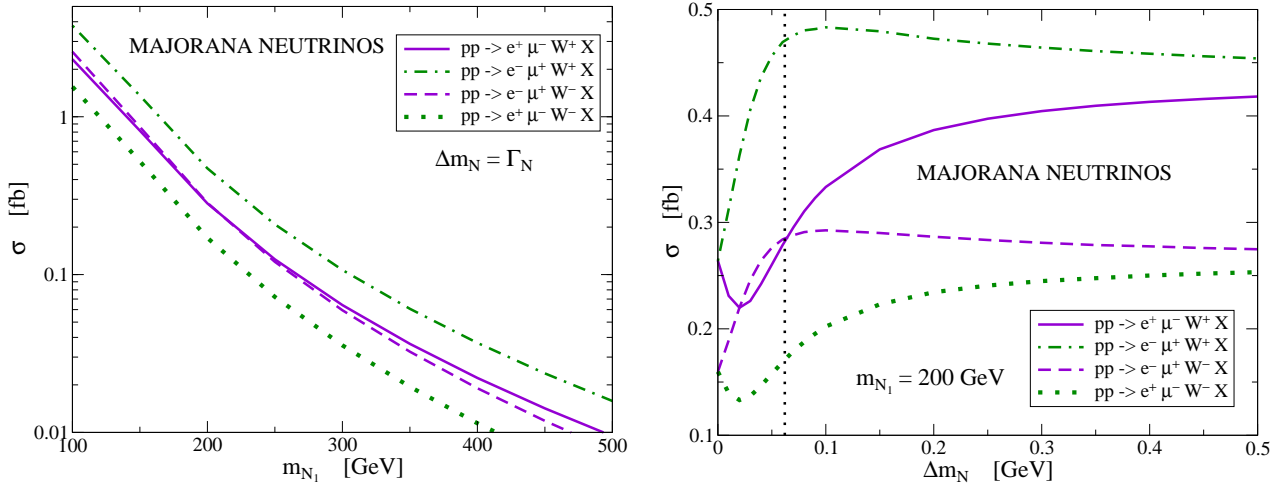

    \begin{center}
      \includegraphics[scale=0.36]{LNC2a}\hspace{0.3cm}
      \includegraphics[scale=0.36]{LNC2b}
    \end{center}
    \caption{\it Cross sections for LNC signals in scenario {\bf
    (S2)}. In this scenario, no CP violation is present for Dirac
    neutrinos, so plots are not shown for these. In this and all other
    plots, the vertical dotted line represents the value of the heavy
    neutrino widths.}
    \label{S2}
    \vspace{0.4cm}
  \end{figure}
  \begin{figure}[ht!]
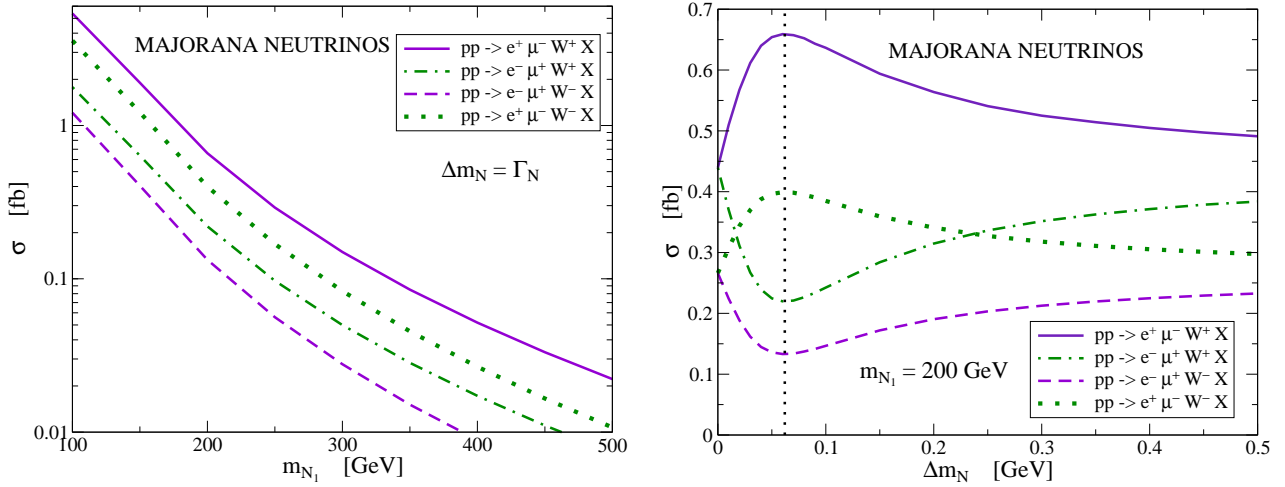

    \begin{center}
      \includegraphics[scale=0.36]{LNC3a}\hspace{0.3cm}
      \includegraphics[scale=0.36]{LNC3b}
    \end{center}
    \caption{\it Cross sections for LNC signals in scenario {\bf (S3)}
    with Majorana neutrinos.}
    \label{S3}
  \end{figure}
  \begin{figure}[ht!]
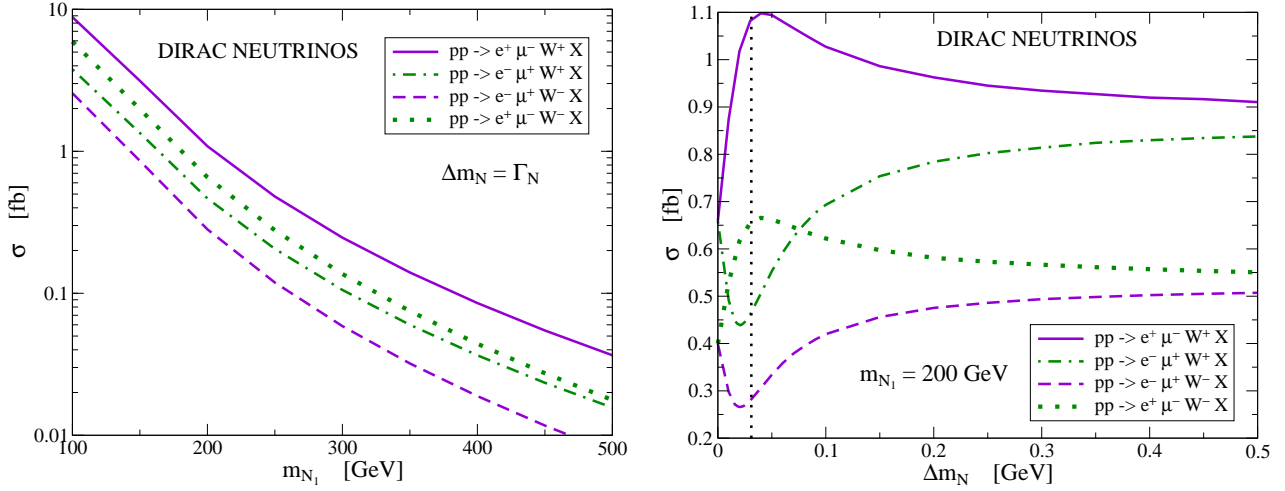

    \begin{center}
      \includegraphics[scale=0.36]{DIRAC3a}\hspace{0.3cm}
      \includegraphics[scale=0.36]{DIRAC3b}
    \end{center}
    \caption{\it As Fig.~\ref{S3}, but with Dirac neutrinos.}
    \label{S6}
    \vspace{0.25cm}
  \end{figure}
  \begin{figure}[ht!]
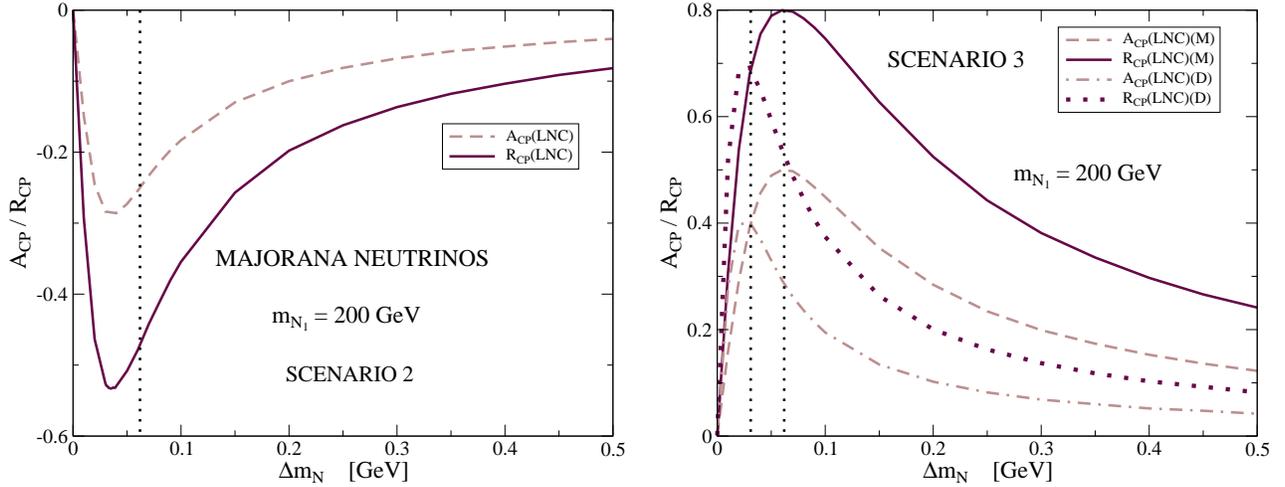

    \begin{center}
      \includegraphics[scale=0.36]{ACPLNC2}\hspace{0.3cm}
      \includegraphics[scale=0.36]{ACPLNC3}
    \end{center}
    \caption{\it The CP-violating observables $A_{\rm CP}({\rm LNC})$
    and $R_{\rm CP}({\rm LNC})$ as defined in (\ref{eq31}) and
    (\ref{eq34}).}
    \label{ACP}
  \end{figure}
  We consider four distinct LNC signals, these are:
  \begin{equation}
    (1)\,pp\to e^+\mu^-W^+X;\hspace{0.5cm}
    (2)\,pp\to e^-\mu^+W^-X;\hspace{0.5cm}
    (3)\,pp\to e^+\mu^-W^-X;\hspace{0.5cm}
    (4)\,pp\to e^-\mu^+W^+X.
    \label{LNCsigs}
  \end{equation}
Signals (1) and  (2) are CP-conjugates of each  other and likewise (3)
and (4). Furthermore,  the cross sections for (3)  and (4) are related
to those for (1) and (2) through
  \begin{equation}
    \sigma(pp\to e^\pm \mu^\mp W^+ X)=K\sigma(pp\to e^\pm \mu^\mp W^- X)\,,
    \label{eq33}
  \end{equation}
which  holds even  when  CP  is not  conserved.  Using this  relation,
$A_{\rm  CP}$ as  defined  in (\ref{eq22})  can  be expressed  without
involving $K$, viz
  \begin{equation}
    A_{\rm CP}({\rm LNC})=\frac{\sigma(pp\to e^+\mu^-W^\pm X)-\sigma(pp\to e^-\mu^+W^\pm X)}{\sigma(pp\to e^+\mu^-W^\pm X)+\sigma(pp\to e^-\mu^+W^\pm X)}\,.
    \label{eq31}
  \end{equation}
This has  also the advantage that  it is not  necessary to distinguish
between the two $W$  boson charges experimentally. $R_{\rm CP}$, given
in (\ref{eq23}), can also  be re-expressed using (\ref{eq33}), and can
thus be written as
  \begin{equation}
    R_{\rm CP}({\rm LNC})=\frac{\displaystyle{\frac{\sigma(pp\to e^+\mu^-W^\pm X)}{\sigma(pp\to e^-\mu^+W^\pm X)}-\frac{\sigma(pp\to e^-\mu^+W^\pm X)}{\sigma(pp\to e^+\mu^-W^\pm X)}}}{\displaystyle{\frac{\sigma(pp\to e^+\mu^-W^\pm X)}{\sigma(pp\to e^-\mu^+W^\pm X)}+\frac{\sigma(pp\to e^-\mu^+W^\pm X)}{\sigma(pp\to e^+\mu^-W^\pm X)}}}\,,
    \label{eq34}
  \end{equation}
where again,  the two charges of  $W$ boson are  summed over. However,
the observable $R_{\rm CP}({\rm LNC})$ turns out to be closely related
to $A_{\rm CP}({\rm LNC})$.

 The cross sections for  the signal processes given in (\ref{LNCsigs})
 are shown in Fig.~\ref{S2} for scenario {\bf (S2)} and Figs. \ref{S3}
 and \ref{S6} for scenario {\bf  (S3)}. Since scenario {\bf (S2)} only
 results in CP violation for Majorana neutrinos, the results for Dirac
 neutrinos are  not shown. We've  plotted the signals as  functions of
 $m_{N_1}$ and  $\Delta m_N$. In  the latter plots, the  point $\Delta
 m_N=\Gamma_N$  is marked  by a  vertical dotted  line\footnote{In the
 scenarios  considered, the  widths  of the  two  heavy neutrinos  are
 almost  equal, as  all couplings  have the  same magnitude.}.  The CP
 asymmetries are close to their maxima at this point, which we use for
 our plots  against $m_{N_1}$.  The CP-violating  observables given in
 (\ref{eq31}) and (\ref{eq34}) are  shown as functions of $\Delta m_N$
 in Fig.~\ref{ACP}.   Again the point $\Delta  m_N=\Gamma_N$ is marked
 by vertical dotted lines (one each for Dirac and Majorana neutrinos).
   
  \subsection{LNV processes}\label{LNV}
  We now analyse the LNV processes:
  \begin{equation}
    (1)\,pp\to e^+e^+W^-X;\hspace{0.5cm}
    (2)\,pp\to e^-e^-W^+X;\hspace{0.5cm}
    (3)\,pp\to e^+\mu^+W^-X;\hspace{0.5cm}
    (4)\,pp\to e^-\mu^-W^+X.
    \label{LNVsigs}
  \end{equation}
We have not included the  di-muon channel processes here since for the
two  scenarios  {\bf(S2)} and  {\bf(S3)},  there  is  no CP  asymmetry
between   these.     However,   interchanging   $B_{e2}\leftrightarrow
B_{\mu2}$  in these scenarios  would give  asymmetries in  the di-muon
channel but not the di-electron channel, which is just as likely.
  \begin{figure}[ht]
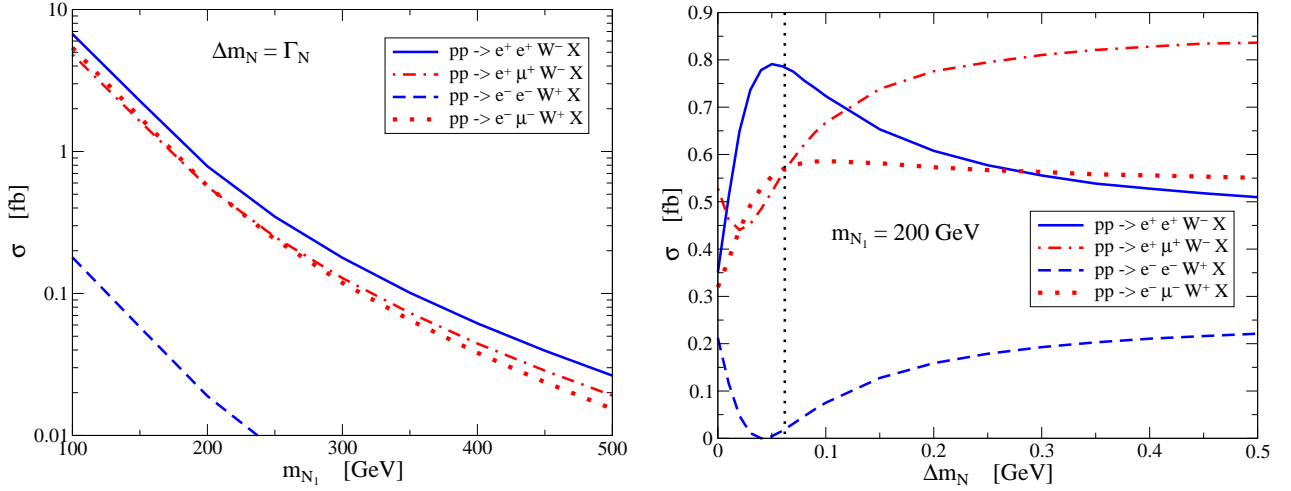

    \vspace{0.28cm}
    \begin{center}
      \includegraphics[scale=0.36]{LNV2a}\hspace{0.3cm}
      \includegraphics[scale=0.36]{LNV2b}
    \end{center}
    \caption{\it Cross sections for LNV signals in scenario {\bf (S2)}.}
    \label{S4}
  \end{figure}
  \begin{figure}[ht]
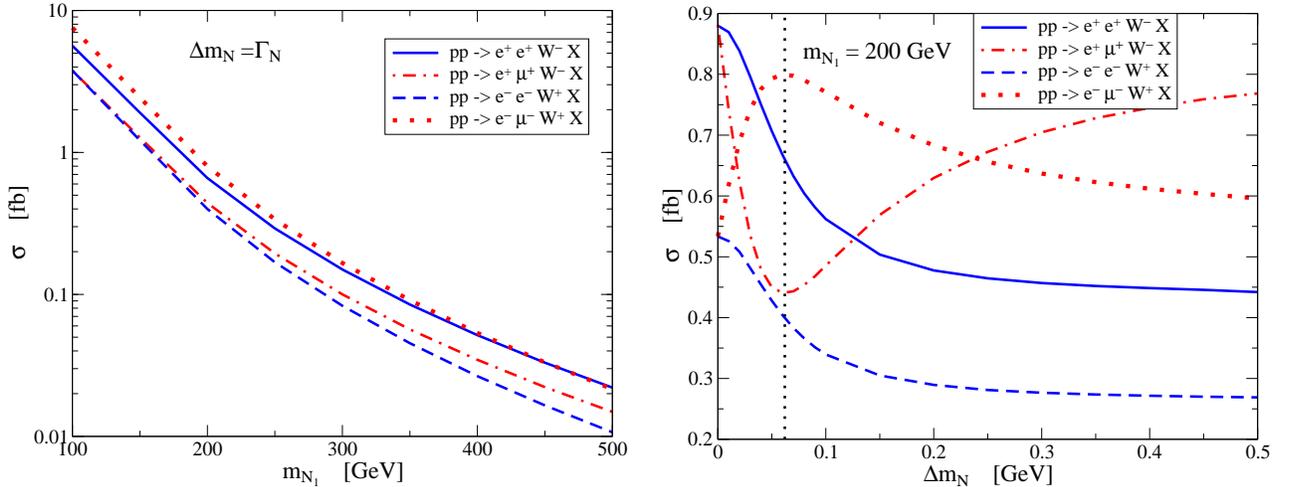

    \begin{center}
      \includegraphics[scale=0.36]{LNV3a}\hspace{0.3cm}
      \includegraphics[scale=0.36]{LNV3b}
    \end{center}
    \caption{\it Cross sections for LNV signals in scenario {\bf
    (S3)}. CP asymmetries are only present for this scenario between
    signals with different flavour final state charged leptons.}
    \label{S5}
    \vspace{0.25cm}
  \end{figure}
  \begin{figure}[ht]
    \begin{center}
      \includegraphics[scale=0.36]{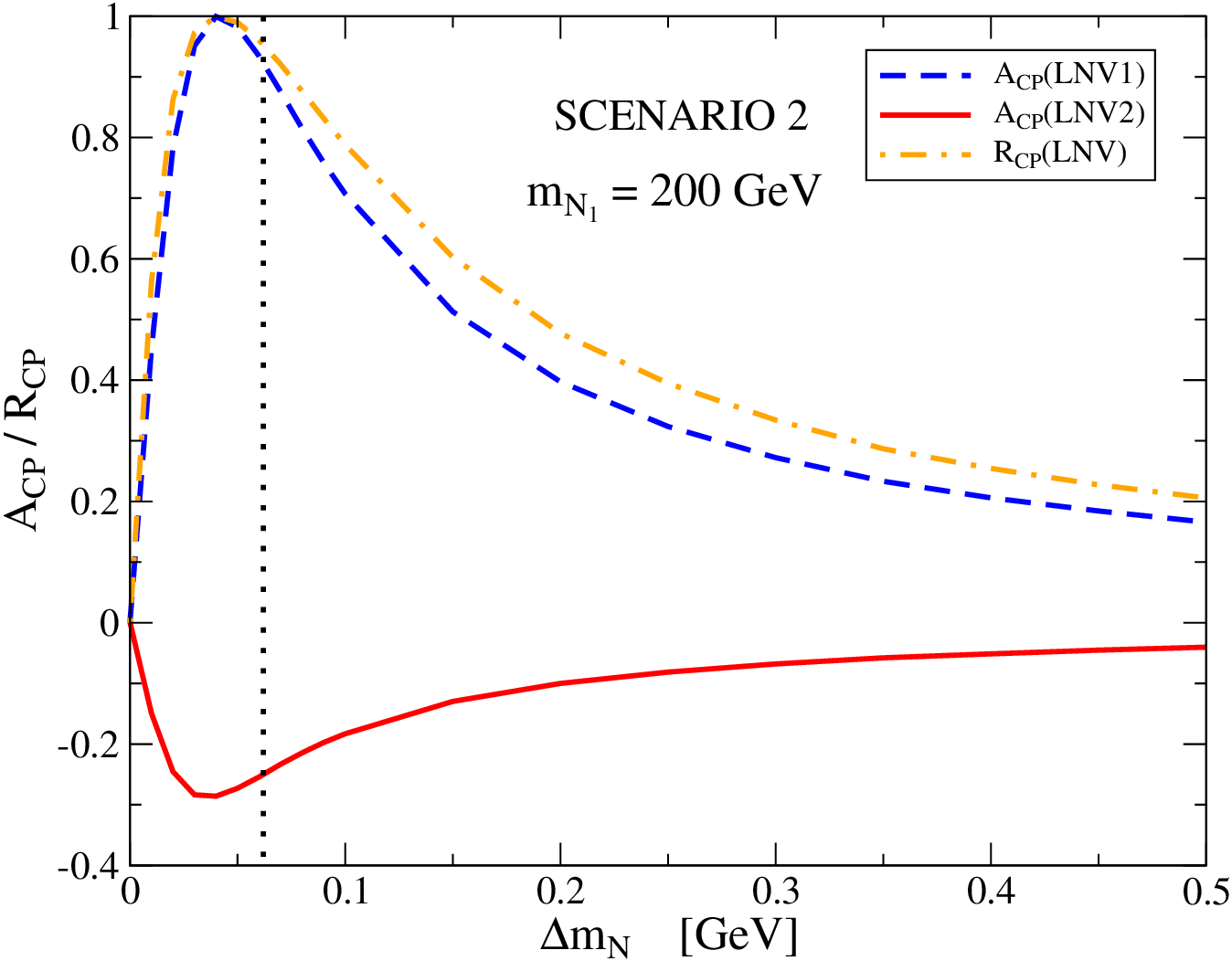}\hspace{0.3cm}
      \includegraphics[scale=0.36]{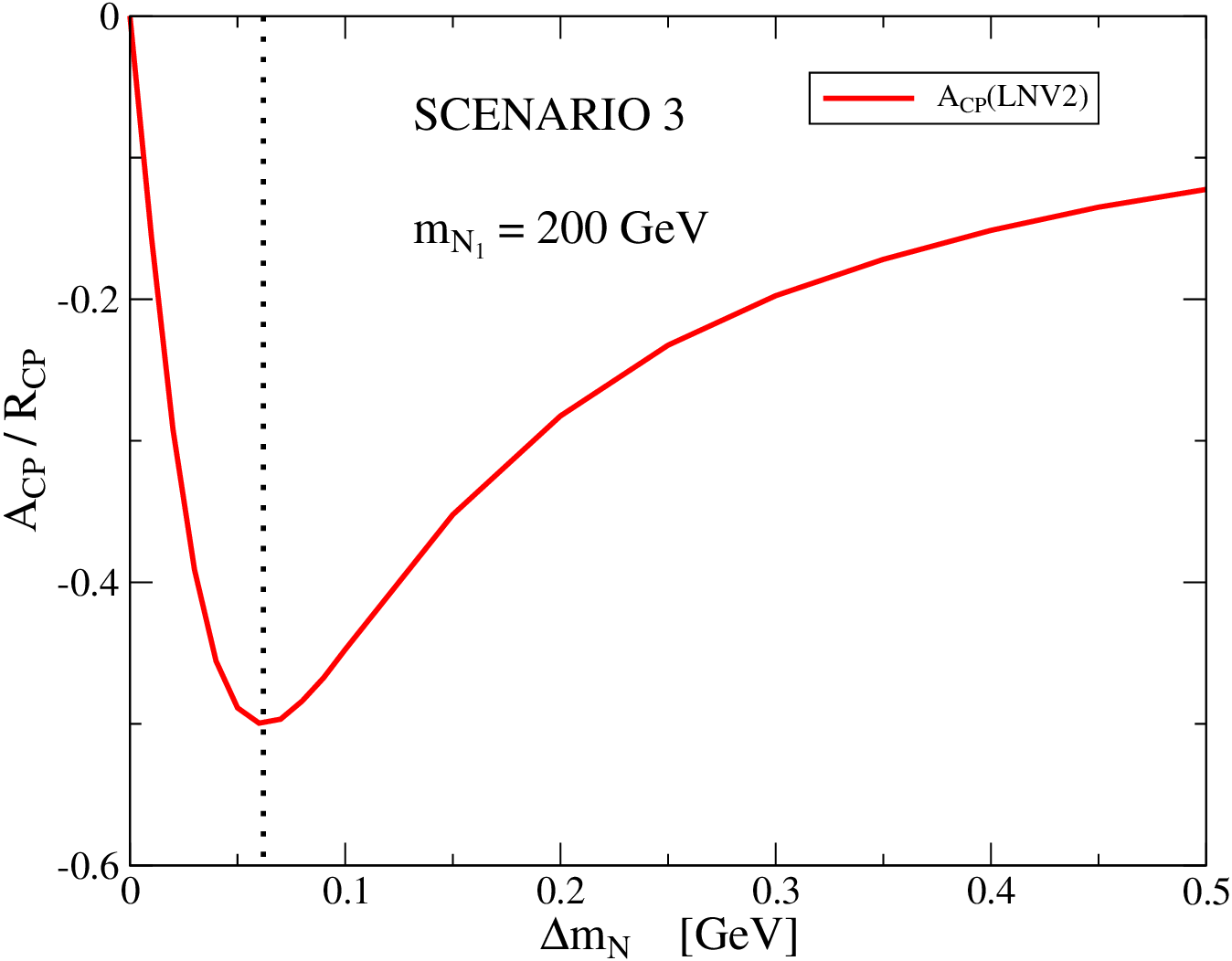}
    \end{center}
    \caption{\it The CP-violating observables $A_{\rm CP}(\rm LNV1)$,
    $A_{\rm CP}(\rm LNV2)$ and $R_{\rm CP}({\rm LNV})$ as defined in
    (\ref{eq35})--(\ref{eq25}). In scenario {\bf (S3)}, $A_{\rm
    CP}(\rm LNV1)=0$ and so $R_{\rm CP}({\rm LNV})=-A_{\rm CP}(\rm
    LNV2)$.}
    \label{ACP2}
  \end{figure}
  \begin{table}[ht]
    \caption{\it Properties of the cross sections and CP asymmetries
    considered in Section \ref{results}}
    \begin{center}
      \begin{tabular}{|c|c|c|c|c|}
	\hline
	Observable(s)&Heavy Neutrino&CP violation&CP violation&$K$-dependence\\
	&Type&in {\bf (S2)}&in {\bf (S3)}&\\
	\hline\hline
	$\sigma(pp\to e^\pm\mu^\mp W^\pm X)$&Dirac&No&Yes&No\\\hline
	$\sigma(pp\to e^\pm\mu^\mp W^\pm X)$&Majorana&Yes&Yes&No\\\hline
	$\sigma(pp\to e^\pm\mu^\mp W^\mp X)$&Dirac&No&Yes&No\\\hline
	$\sigma(pp\to e^\pm\mu^\mp W^\mp X)$&Majorana&Yes&Yes&No\\\hline
	$\sigma(pp\to e^\pm e^\pm W^\mp X)$&Majorana&Yes&No&No\\\hline
	$\sigma(pp\to\mu^\pm\mu^\pm W^\mp X)$&Majorana&No&No&No\\\hline
	$\sigma(pp\to e^\pm\mu^\pm W^\mp X)$&Majorana&Yes&Yes&No\\\hline
	$A_{\rm CP}({\rm LNC})$&Dirac&No&Yes&No\\\hline
	$A_{\rm CP}({\rm LNC})$&Majorana&Yes&Yes&No\\\hline
	$R_{\rm CP}({\rm LNC})$&Dirac&No&Yes&No\\\hline
	$R_{\rm CP}({\rm LNC})$&Majorana&Yes&Yes&No\\\hline
	$A_{\rm CP}(\rm LNV1)$&Majorana&Yes&No&Yes\\\hline
	$A_{\rm CP}(\rm LNV2)$&Majorana&Yes&Yes&Yes\\\hline
	$R_{\rm CP}({\rm LNV})$&Majorana&Yes&Yes&No\\\hline
      \end{tabular}
    \end{center}
    \label{Tab}
  \end{table}

As before, $A_{\rm CP}$ and $R_{\rm CP}$ can be defined as follows:
  \begin{eqnarray}
    A_{\rm CP}(\rm LNV1)&=&\frac{\sigma(pp\to e^+e^+W^- X)-K\sigma(pp\to e^-e^-W^+ X)}{\sigma(pp\to e^+e^+W^- X)+K\sigma(pp\to e^-e^-W^+ X)}\,,\label{eq35}\\
    A_{\rm CP}(\rm LNV2)&=&\frac{\sigma(pp\to e^+\mu^+W^- X)-K\sigma(pp\to e^-\mu^-W^+ X)}{\sigma(pp\to e^+\mu^+W^- X)+K\sigma(pp\to e^-\mu^-W^+ X)}\,,\\
    R_{\rm CP}({\rm LNV})&=&\frac{\displaystyle{\frac{\sigma(pp\to e^+e^+W^- X)}{\sigma(pp\to e^+\mu^+W^- X)}-\frac{\sigma(pp\to e^-e^-W^+ X)}{\sigma(pp\to e^-\mu^-W^+ X)}}}{\displaystyle{\frac{\sigma(pp\to e^+e^+W^- X)}{\sigma(pp\to e^+\mu^+W^- X)}+\frac{\sigma(pp\to e^-e^-W^+ X)}{\sigma(pp\to e^-\mu^-W^+ X)}}}\label{eq25}\,.
  \end{eqnarray}
The signal cross sections given  in (\ref{LNVsigs}) are plotted in the
same manner as the LNC  signals were. These are shown in Fig.~\ref{S4}
for  scenario   {\bf  (S2)}   and  Fig.~\ref{S5}  for   scenario  {\bf
(S3)}.    Similarly,   the    CP-violating   observables    given   in
(\ref{eq35})--(\ref{eq25})  are shown  in  Fig.~\ref{ACP2}. The  cross
sections  for both  the LNC  and LNV  signals fall  very  rapidly with
increasing heavy neutrino mass.  For $|B_{lN}|=0.05$ and an integrated
luminosity of 100  fb$^{-1}$, to produce at least  10 signal events in
any single  channel would typically  require $m_N\lesssim300-400\,{\rm
GeV}$. This could be enough  to not just discover heavy neutrinos, but
also observe  resonant CP violation as  well. As can be  seen from the
plots  in  Figs.  \ref{ACP}  and  \ref{ACP2}, the  CP  asymmetries  in
scenarios  {\bf  (S2)} and  {\bf  (S3)}  are  very large  for  $\Delta
m_N\approx\Gamma_N$.     With    this     condition,     and    taking
$m_{N_1}=300\,{\rm GeV}$ for example, scenario {\bf (S2)} would result
in  about 20  signal events  for $pp\to  e^+e^+W^-X$ without  a single
$pp\to e^-e^-W^+X$ event likely to be observed.

In  the limit  $\Delta m_N=0$,  the  asymmetries vanish.   This is  as
expected from  Section \ref{large}, since  there it is shown  that for
couplings of equal magnitude,  the asymmetries are proportional to the
mass-splitting  of  the  heavy  neutrinos. For  the  two  CP-violating
scenarios  considered,  the CP  asymmetries  are  larger for  Majorana
neutrinos than for  Dirac ones.  However, more examples  would need to
be calculated to determine if this is a general trend.  Since they are
constructed out  of the asymmetries between  two independent processes
and  their CP-conjugates,  it  should  come as  no  surprise that  the
$R_{\rm CP}$ asymmetries  are larger than the $A_{\rm  CP}$ ones.  The
one  exception to  this is  $R_{\rm CP}({\rm  LNV})$ in  scenario {\bf
(S3)}.  This is  because there is no asymmetry between  one of the two
pairs of  CP-conjugate processes it  is constructed from.  For $\Delta
m_N=\Gamma_N$, both  $A_{\rm CP}$ and $R_{\rm CP}$  are independent of
the heavy neutrino  mass scale, although this is  not obvious from our
plots since $K$ does depend on $m_N$.

\section{Conclusions}\label{conclusion}

Based     on     the     resummation    formalism     developed     in
\cite{Pilaftsis:1997dr,Pilaftsis:1997jf},   we  have   calculated  the
$2\times2$  heavy  neutrino  propagator  matrix  for  both  Dirac  and
Majorana neutrinos. Our  results apply to scenarios for  which the two
heavy  neutrinos  that are  nearly  degenerate  in  mass dominate  the
production cross sections. The  formalism involves the absorptive part
of the heavy  neutrino self-energy, which is given  to one-loop, again
for both Dirac and Majorana neutrinos.

Assuming a two  heavy neutrino mixing system, we  have given numerical
estimates  for the  production cross  sections  of LFV  and LNV  heavy
neutrino signal  processes at  the LHC. These  are of the  form $pp\to
ll'WX$, with the final state  particle charges (not including the beam
remnants $X$)  adding up to  $\pm1$. For Dirac neutrinos,  the leptons
have  to  have  opposite charges  and  so,  in  order to  avoid  large
backgrounds, $l$ and $l'$ should be different. For Majorana neutrinos,
LNV signals of the form $pp\to l^\pm l'^\pm W^\mp X$ are also allowed,
where  $l$  and $l'$  can  (but  do not  have  to)  be  equal. The  SM
background to  either of these  types of signals should  be negligible
after cuts. For the magnitudes of couplings we used ($|B_{lN}|=0.05$),
heavy  neutrinos will be  observable at  the LHC  if they  have masses
between  about $100\,{\rm  GeV}$ and  about $400\,{\rm  GeV}$.  If the
heavy neutrinos are  much lighter than $100\,{\rm GeV}$,  the LEP data
put severe limits on the $B$-couplings, leading to unobservable signal
cross sections.

We have plotted the signal  cross sections in three different possible
scenarios for the couplings of  the heavy neutrinos. In the first, all
the  couplings are  real. This  is  the CP-conserving  limit which  is
needed  to calculate  the asymmetry  due  to the  different PDF's  for
producing $W^+$ and  $W^-$ bosons. The second and  third scenarios are
chosen as examples in which large  CP asymmetries exist in a number of
channels. CP-violating  observables constructed from  the signal cross
sections are also  plotted. For a mass-splitting of  the two neutrinos
of order  their widths,  these asymmetries can  be very large  or even
maximal, giving rise to resonant CP violation. It should be noted that
these  scenarios require  additional flavour  symmetries in  the heavy
neutrino sector in order  to be possible. Experimental constraints, in
particular  those from the  lack of  observation of  $\mu\to e\gamma$,
require contributions from additional particles of new physics (either
further  heavy neutrinos  or something  else) to  be  satisfied. These
extra particles would  need to cause quite large  cancellations in the
rate of this process without  making a significant contribution to the
signals  of   interest.  Also,  for  Majorana   neutrinos,  there  are
theoretical  constraints that  require at  least two  additional heavy
neutrinos (four in  total) to be present in the theory  in order to be
satisfied. Nevertheless,  our results  demonstrate that very  large CP
asymmetries are possible at the  LHC. The couplings used have not been
motivated in any  fashion, but neither are they  unique. In particular
it should be kept in mind that an interchange of charged lepton labels
gives scenarios which are no more or less likely.

Among the  different processes  we have been  studying here,  the most
realistic modes  to look for large  CP asymmetries at the  LHC are the
di-muon or di-electron channels. These processes are LNV and hence are
only  allowed  for  Majorana  neutrinos.  For these  channels,  it  is
possible to have  the heavy neutrino couplings to  either the electron
or  muon   heavily  suppressed,  hence   satisfying  the  experimental
constraints. It may then be  possible to have resonant CP violation in
these channels  without resorting to additional  flavour symmetries or
cancellations.

The analysis presented  here could be extended to  other colliders, in
particular to the ILC. Here the signals $e^+e^-\to l^\pm W^\mp\nu$ can
be considered which are CP-conjugates of each other. The ILC should be
a cleaner  experimental environment and  depending on its  cms energy,
may well  be able  to produce heavy  neutrinos of  considerably larger
mass. However, the ILC would suffer  compared to the LHC in the search
for  heavy neutrinos in  the sense  that the  SM background  cannot so
easily be  suppressed. Linked to  this, evidence of $L$  violation and
hence whether or not neutrinos are Majorana particles is far harder to
obtain. Also, an observable  signal requires a significant coupling of
the heavy  neutrino to the electron,  a requirement not  shared by the
LHC.

In  summary,  resonant CP  violation  due  to electroweak-scale  heavy
neutrinos is  an interesting possibility  that might well  be observed
for the first time at the  LHC. If realised, this could, in principle,
give  an explanation for  the BAU  through resonant  leptogenesis. The
observation  of  electroweak-scale heavy  neutrinos  may not  directly
unravel the detailed structure of  the light neutrino mass matrix, but
it will  naturally point towards  scenarios based on  some approximate
lepton-number  or flavour symmetry  (see also  our comment  in Section
2.1).  This  approximate lepton-number symmetry may  be imprinted into
the relative  decay rates  of the heavy  neutrinos into  the different
charged-lepton flavours $e$, $\mu$  and $\tau$, from which an estimate
of  the  large elements  of  the Dirac  mass  matrix  $m_D$ could,  in
principle, be  obtained.  Possible new signatures  or constraints from
low-energy experiments, including  the low-energy neutrino oscillation
data, may offer valuable information to restrict the form of $m_D$ and
the  light-to-heavy  neutrino-mixing  matrix  $m_D/m_M$.   A  detailed
analysis of possible correlations between resonant CP violation due to
heavy neutrinos which we have been studying here and other observables
may be the subject of a future communication.

\section*{Acknowledgements}
We  thank  Mrinal  Dasgupta  and  Jeff Forshaw  for  a  discussion  of
hadronization  and colour  interference effects.   We also  thank Shun
Zhou  for pointing  us out  a  typo in~(\ref{Amatrix}).   The work  of
S.B.~has been funded by the PPARC studentship PPA/S/S/2003/03666.  The
work of J.S.L.~was supported in  part by the Korea Research Foundation
(KRF) and  the Korean Federation  of Science and  Technology Societies
Grant and in part by  the KRF grant: KRF-2005-084-C00001 funded by the
Korea Government (MOEHRD, Basic Research Promotion Fund).  The work of
A.P.~was  supported in  part by  the PPARC  grants:  PP/D0000157/1 and
PP/C504286/1.

\vspace{1cm}

\section*{Note Added} Shortly after communicating our paper, we became
aware of  a new analysis of  the background contributing  to the heavy
neutrino signals~\cite{delAguila:2007em}.  According to this analysis,
previous  studies  have grossly  underestimated  the  background by  a
factor of $\sim 30$, because they did not take into account high-order
pile-up processes of jets which cannot be reduced by various kinematic
cuts.   As a  consequence, those  authors find  that the  LHC  will be
capable of only probing relatively  light heavy neutrinos of masses up
to  175~GeV, thus restricting  the results  of our  study accordingly.
Nonetheless,  it  should  be  noticed  that  the  background  of  high
multiplicity jet processes stems predominantly from colour non-singlet
states, unlike the 2 distinct  jets in the signal which originate from
decays of  the colourless heavy  neutrinos or $W^\pm$  bosons.  Hence,
the signal and  background processes are expected to  show a different
topology of hadronic activities in  the central region, which could be
used to eliminate the contribution of the high-order pile-up processes
of  jets, with the  hope to  extend the  reach of  the LHC  to heavier
Majorana neutrinos.   Such an analysis  lies beyond the scope  of this
paper.

\newpage

  \bibliography{Paper}

\begin{mcbibliography}{10}

\bibitem{Fukuda:1998mi}
Super-Kamiokande, Y.~Fukuda {\em et~al.},
\newblock Phys. Rev. Lett. {\bf 81}, 1562 (1998), hep-ex/9807003\relax
\relax
\bibitem{Apollonio:1999ae}
CHOOZ, M.~Apollonio {\em et~al.},
\newblock Phys. Lett. {\bf B466}, 415 (1999), hep-ex/9907037\relax
\relax
\bibitem{Ahmad:2002jz}
SNO, Q.~R. Ahmad {\em et~al.},
\newblock Phys. Rev. Lett. {\bf 89}, 011301 (2002), nucl-ex/0204008\relax
\relax
\bibitem{Ahn:2002up}
K2K, M.~H. Ahn {\em et~al.},
\newblock Phys. Rev. Lett. {\bf 90}, 041801 (2003), hep-ex/0212007\relax
\relax
\bibitem{Eguchi:2002dm}
KamLAND, K.~Eguchi {\em et~al.},
\newblock Phys. Rev. Lett. {\bf 90}, 021802 (2003), hep-ex/0212021\relax
\relax
\bibitem{Achard:2001qv}
L3, P.~Achard {\em et~al.},
\newblock Phys. Lett. {\bf B517}, 67 (2001), hep-ex/0107014\relax
\relax
\bibitem{Pilaftsis:1991ug}
A.~Pilaftsis,
\newblock Z. Phys. {\bf C55}, 275 (1992), hep-ph/9901206\relax
\relax
\bibitem{Datta:1993nm}
A.~Datta, M.~Guchait, and A.~Pilaftsis,
\newblock Phys. Rev. {\bf D50}, 3195 (1994), hep-ph/9311257\relax
\relax
\bibitem{Almeida:2000pz}
J.~Almeida, F. M.~L., Y.~A. Coutinho, J.~A. Martins~Simoes, and M.~A.~B.
  do~Vale,
\newblock Phys. Rev. {\bf D62}, 075004 (2000), hep-ph/0002024\relax
\relax
\bibitem{Panella:2001wq}
O.~Panella, M.~Cannoni, C.~Carimalo, and Y.~N. Srivastava,
\newblock Phys. Rev. {\bf D65}, 035005 (2002), hep-ph/0107308\relax
\relax
\bibitem{Han:2006ip}
T.~Han and B.~Zhang,
\newblock Phys. Rev. Lett. {\bf 97}, 171804 (2006), hep-ph/0604064\relax
\relax
\bibitem{delAguila:2006dx}
F.~del Aguila, J.~A. Aguilar-Saavedra, and R.~Pittau,
\newblock J. Phys. Conf. Ser. {\bf 53}, 506 (2006), hep-ph/0606198\relax
\relax
\bibitem{Gluza:1996bz}
J.~Gluza and M.~Zralek,
\newblock Phys. Rev. {\bf D55}, 7030 (1997), hep-ph/9612227\relax
\relax
\bibitem{Cvetic:1998vg}
G.~Cvetic, C.~S. Kim, and C.~W. Kim,
\newblock Phys. Rev. Lett. {\bf 82}, 4761 (1999), hep-ph/9812525\relax
\relax
\bibitem{Almeida:2000yx}
J.~Almeida, F. M.~L., Y.~A. Coutinho, J.~A. Martins~Simoes, and M.~A.~B.
  do~Vale,
\newblock Phys. Rev. {\bf D63}, 075005 (2001), hep-ph/0008201\relax
\relax
\bibitem{delAguila:2005mf}
F.~del Aguila, J.~A. Aguilar-Saavedra, A.~Martinez de~la Ossa, and D.~Meloni,
\newblock Phys. Lett. {\bf B613}, 170 (2005), hep-ph/0502189\relax
\relax
\bibitem{Peressutti:2001ms}
J.~Peressutti, O.~A. Sampayo, and J.~I. Aranda,
\newblock Phys. Rev. {\bf D64}, 073007 (2001), hep-ph/0105162\relax
\relax
\bibitem{Bray:2005wv}
S.~Bray, J.~S. Lee, and A.~Pilaftsis,
\newblock Phys. Lett. {\bf B628}, 250 (2005), hep-ph/0508077\relax
\relax
\bibitem{Flanz:1994yx}
M.~Flanz, E.~A. Paschos, and U.~Sarkar,
\newblock Phys. Lett. {\bf B345}, 248 (1995), hep-ph/9411366\relax
\relax
\bibitem{Flanz:1996fb}
M.~Flanz, E.~A. Paschos, U.~Sarkar, and J.~Weiss,
\newblock Phys. Lett. {\bf B389}, 693 (1996), hep-ph/9607310\relax
\relax
\bibitem{Covi:1996wh}
L.~Covi, E.~Roulet, and F.~Vissani,
\newblock Phys. Lett. {\bf B384}, 169 (1996), hep-ph/9605319\relax
\relax
\bibitem{Pilaftsis:1997jf}
A.~Pilaftsis,
\newblock Phys. Rev. {\bf D56}, 5431 (1997), hep-ph/9707235\relax
\relax
\bibitem{Pilaftsis:2003gt}
A.~Pilaftsis and T.~E.~J. Underwood,
\newblock Nucl. Phys. {\bf B692}, 303 (2004), hep-ph/0309342\relax
\relax
\bibitem{Pilaftsis:2005rv}
A.~Pilaftsis and T.~E.~J. Underwood,
\newblock Phys. Rev. {\bf D72}, 113001 (2005), hep-ph/0506107\relax
\relax
\bibitem{Garbrecht:2006az}
B.~Garbrecht, C.~Pallis, and A.~Pilaftsis,
\newblock JHEP {\bf 12}, 038 (2006), hep-ph/0605264\relax
\relax
\bibitem{Branco:2006hz}
G.~C. Branco, A.~J. Buras, S.~Jager, S.~Uhlig, and A.~Weiler,
\newblock (2006), hep-ph/0609067\relax
\relax
\bibitem{Pumplin:2002vw}
J.~Pumplin {\em et~al.},
\newblock JHEP {\bf 07}, 012 (2002), hep-ph/0201195\relax
\relax
\bibitem{Martin:2002aw}
A.~D. Martin, R.~G. Roberts, W.~J. Stirling, and R.~S. Thorne,
\newblock Eur. Phys. J. {\bf C28}, 455 (2003), hep-ph/0211080\relax
\relax
\bibitem{Pilaftsis:1997dr}
A.~Pilaftsis,
\newblock Nucl. Phys. {\bf B504}, 61 (1997), hep-ph/9702393\relax
\relax
\bibitem{Fritzsch:1974nn}
H.~Fritzsch and P.~Minkowski,
\newblock Ann. Phys. {\bf 93}, 193 (1975)\relax
\relax
\bibitem{Minkowski:1977sc}
P.~Minkowski,
\newblock Phys. Lett. {\bf B67}, 421 (1977)\relax
\relax
\bibitem{Gell-Mann:1979}
M.~Gell-Mann, P.~Ramond, and R.~Slansky,
\newblock in {\em Supergravity}, edited by P.~van Nieuwenhuizen and
  D.~Friedman, p. 315, North-Holland, Amsterdam, 1979\relax
\relax
\bibitem{Yanagida:1979}
T.~Yanagida,
\newblock in {\em Preceedings of the Workshop on the Unified Theories and
  Baryon Number of the Universe}, edited by O.~Sawada and A.~Sugamoto, KEK,
  Tsukuba, 1979\relax
\relax
\bibitem{Mohapatra:1979ia}
R.~N. Mohapatra and G.~Senjanovic,
\newblock Phys. Rev. Lett. {\bf 44}, 912 (1980)\relax
\relax
\bibitem{Ellis:2006mg}
J.~Ellis, M.~E. Gomez, and S.~Lola,
\newblock (2006), hep-ph/0612292\relax
\relax
\bibitem{Witten:1985bz}
E.~Witten,
\newblock Nucl. Phys. {\bf B268}, 79 (1986)\relax
\relax
\bibitem{Mohapatra:1986bd}
R.~N. Mohapatra and J.~W.~F. Valle,
\newblock Phys. Rev. {\bf D34}, 1642 (1986)\relax
\relax
\bibitem{Gluza:2002vs}
J.~Gluza,
\newblock Acta Phys. Polon. {\bf B33}, 1735 (2002), hep-ph/0201002\relax
\relax
\bibitem{Altarelli:2004za}
G.~Altarelli and F.~Feruglio,
\newblock New J. Phys. {\bf 6}, 106 (2004), hep-ph/0405048\relax
\relax
\bibitem{Wyler:1982dd}
D.~Wyler and L.~Wolfenstein,
\newblock Nucl. Phys. {\bf B218}, 205 (1983)\relax
\relax
\bibitem{Nandi:1985uh}
S.~Nandi and U.~Sarkar,
\newblock Phys. Rev. Lett. {\bf 56}, 564 (1986)\relax
\relax
\bibitem{Valle:1990pk}
J.~W.~F. Valle,
\newblock Prog. Part. Nucl. Phys. {\bf 26}, 91 (1991)\relax
\relax
\bibitem{Langacker:1988ur}
P.~Langacker and D.~London,
\newblock Phys. Rev. {\bf D38}, 886 (1988)\relax
\relax
\bibitem{Pontecorvo:1957cp}
B.~Pontecorvo,
\newblock Sov. Phys. JETP {\bf 6}, 429 (1957)\relax
\relax
\bibitem{Pontecorvo:1957qd}
B.~Pontecorvo,
\newblock Sov. Phys. JETP {\bf 7}, 172 (1958)\relax
\relax
\bibitem{Maki:1962mu}
Z.~Maki, M.~Nakagawa, and S.~Sakata,
\newblock Prog. Theor. Phys. {\bf 28}, 870 (1962)\relax
\relax
\bibitem{Cheng:1980tp}
T.~P. Cheng and L.-F. Li,
\newblock Phys. Rev. Lett. {\bf 45}, 1908 (1980)\relax
\relax
\bibitem{Korner:1992an}
J.~G. Korner, A.~Pilaftsis, and K.~Schilcher,
\newblock Phys. Lett. {\bf B300}, 381 (1993), hep-ph/9301290\relax
\relax
\bibitem{Bernabeu:1993up}
J.~Bernabeu, J.~G. Korner, A.~Pilaftsis, and K.~Schilcher,
\newblock Phys. Rev. Lett. {\bf 71}, 2695 (1993), hep-ph/9307295\relax
\relax
\bibitem{Burgess:1993vc}
C.~P. Burgess, S.~Godfrey, H.~Konig, D.~London, and I.~Maksymyk,
\newblock Phys. Rev. {\bf D49}, 6115 (1994), hep-ph/9312291\relax
\relax
\bibitem{Nardi:1994iv}
E.~Nardi, E.~Roulet, and D.~Tommasini,
\newblock Phys. Lett. {\bf B327}, 319 (1994), hep-ph/9402224\relax
\relax
\bibitem{Bhattacharya:1994bj}
G.~Bhattacharya, P.~Kalyniak, and I.~Melo,
\newblock Phys. Rev. {\bf D51}, 3569 (1995), hep-ph/9503248\relax
\relax
\bibitem{Deppisch:2005zm}
F.~Deppisch, T.~S. Kosmas, and J.~W.~F. Valle,
\newblock Nucl. Phys. {\bf B752}, 80 (2006), hep-ph/0512360\relax
\relax
\bibitem{Ilakovac:1994kj}
A.~Ilakovac and A.~Pilaftsis,
\newblock Nucl. Phys. {\bf B437}, 491 (1995), hep-ph/9403398\relax
\relax
\bibitem{Bergmann:1998rg}
S.~Bergmann and A.~Kagan,
\newblock Nucl. Phys. {\bf B538}, 368 (1999), hep-ph/9803305\relax
\relax
\bibitem{Illana:2000ic}
J.~I. Illana and T.~Riemann,
\newblock Phys. Rev. {\bf D63}, 053004 (2001), hep-ph/0010193\relax
\relax
\bibitem{Cvetic:2002jy}
G.~Cvetic, C.~Dib, C.~S. Kim, and J.~D. Kim,
\newblock Phys. Rev. {\bf D66}, 034008 (2002), hep-ph/0202212\relax
\relax
\bibitem{Aubert:2005ye}
BABAR, B.~Aubert {\em et~al.},
\newblock Phys. Rev. Lett. {\bf 95}, 041802 (2005), hep-ex/0502032\relax
\relax
\bibitem{Aubert:2005wa}
BABAR, B.~Aubert {\em et~al.},
\newblock Phys. Rev. Lett. {\bf 96}, 041801 (2006), hep-ex/0508012\relax
\relax
\bibitem{Belanger:1995nh}
G.~Belanger, F.~Boudjema, D.~London, and H.~Nadeau,
\newblock Phys. Rev. {\bf D53}, 6292 (1996), hep-ph/9508317\relax
\relax
\bibitem{Cornwall:1989gv}
J.~M. Cornwall and J.~Papavassiliou,
\newblock Phys. Rev. {\bf D40}, 3474 (1989)\relax
\relax
\bibitem{Papavassiliou:1989zd}
J.~Papavassiliou,
\newblock Phys. Rev. {\bf D41}, 3179 (1990)\relax
\relax
\bibitem{Binosi:2002ft}
D.~Binosi and J.~Papavassiliou,
\newblock Phys. Rev. {\bf D66}, 111901 (2002), hep-ph/0208189\relax
\relax
\bibitem{Binosi:2003rr}
D.~Binosi and J.~Papavassiliou,
\newblock J. Phys. {\bf G30}, 203 (2004), hep-ph/0301096\relax
\relax
\bibitem{Papavassiliou:1995fq}
J.~Papavassiliou and A.~Pilaftsis,
\newblock Phys. Rev. Lett. {\bf 75}, 3060 (1995), hep-ph/9506417\relax
\relax
\bibitem{Papavassiliou:1995gs}
J.~Papavassiliou and A.~Pilaftsis,
\newblock Phys. Rev. {\bf D53}, 2128 (1996), hep-ph/9507246\relax
\relax
\bibitem{Papavassiliou:1996zn}
J.~Papavassiliou and A.~Pilaftsis,
\newblock Phys. Rev. {\bf D54}, 5315 (1996), hep-ph/9605385\relax
\relax
\bibitem{delAguila:2007em}
F.~del Aguila, J.~A. Aguilar-Saavedra, and R.~Pittau,
\newblock (2007), hep-ph/0703261\relax
\relax
\end{mcbibliography}

\end{document}